\documentclass[%
 reprint,
 amsmath,amssymb,
 aps,
 nofootinbib
 prb,
 superscriptaddress
 ]{revtex4-1}
\usepackage{scrextend}
\usepackage{graphicx}
\usepackage{dcolumn}
\usepackage{bm}
\usepackage{tabularx}
\usepackage[utf8]{inputenc}
\usepackage[english]{babel}
\usepackage[bottom]{footmisc}
\usepackage{color}

\usepackage{amsmath}
\usepackage{soul,xcolor}
\setstcolor{red}
\usepackage{hyperref}
\usepackage{booktabs}
\usepackage{epstopdf}

\DeclareMathAlphabet{\pazocal}{OMS}{zplm}{m}{n}

\begin{document}

\preprint{APS/123-QED}

\title{Simultaneous suppression and temperature-driven
crossover of particle-like and wave-like thermal
transport in trigonal CsAg$_3$S$_2$}
\author{Vivek Gusain}
\author{B. K. Mani}
\email{bkmani@physics.iitd.ac.in}
\affiliation {Department of Physics, Indian Institute of Technology,
Hauz Khas, New Delhi 110016, India}

\date{\today}

\begin{abstract}
The simultaneous suppression of particle-like
($\kappa_p$ ) and wave-like ($\kappa_c$)
thermal transport channels in crystalline
materials remains a challenging problem,
as reducing one channel can enhance the 
other. Here, we investigate the thermal
transport in trigonal $P\Bar{3}m1$ 
CsAg$_3$S$_2$ using self-consistent
phonon theory with cubic and 
quartic anharmonicity and the unified
theory of thermal transport.
 Our calculations predict an ultralow lattice
thermal conductivity of 0.222 W m$^{-1}$ K$^{-1}$ 
at 300 K, featuring simultaneously low
$\kappa_p$ and $\kappa_c$ with a weak
temperature dependence of $\sim$T$^{-0.42}$, which
deviates significantly
from the conventional $\sim$T$^{-1}$ behavior. 
This simultaneous suppression arises
from localized Cs modes, which enhance
anharmonic scattering, and the softer
Ag-S framework, which reduces phonon 
group velocities, together suppressing
$\kappa_p$, in conjunction with sparse,
well-separated high-frequency S-derived
optical modes that suppress $\kappa_c$
by limiting the coherent coupling 
between the phonon modes. Furthermore,
the dominant transport channel exhibits
a temperature-dependent crossover, with 
$\kappa_p$ contributing $\sim$74 \% of
total lattice thermal conductivity at
300~K and $\kappa_c$ contributing
$\sim$63 \% at 700~K.  
Combining the ultralow thermal
conductivity with electrical transport
calculated within the momentum relaxation
time approximation (MRTA), we obtain a 
$ZT$ of 1.28 for $n$-type doping at 700 K. 
The present work not only clarifies the
microscopic mechanism of ultralow
thermal conductivity in trigonal
CsAg$_3$S$_2$ but also identifies
the covalent M-Ch framework 
modification within the AM$_3$Ch$_2$
family of ternary chalcogenides as 
an additional tuning parameter for
further reducing $\kappa_p$ within 
the bonding-hierarchy design principle.
These findings motivate further exploration
of framework engineering across broader
classes of low lattice thermal conductivity
materials, featuring a loosely bound rattler
sublattice with a stiff covalent framework.
\end{abstract}

\maketitle

\section{Introduction}
Materials with ultralow thermal conductivity 
are of great importance for various 
technological applications, including 
thermoelectrics \cite{snyder08, mukherjee22}, thermal 
barrier coatings \cite{wang25}, thermal 
insulation \cite{liu22, Zhou23}, and nanoscale thermal 
confinement \cite{du26}. This has motivated
the design and discovery of several classes 
of materials with intrinsically low lattice
thermal conductivity, such as clathrates \cite{takabatake14}, 
filled skutterudites \cite{tang15, li24},
chalcogenides \cite{zhao14, pei20, jia24}, 
Zintl phases \cite{chen19, wei23, zeeshan22}, argyrodites \cite{ren23}.
The suppressed thermal transport in these materials
originates from diverse structural and dynamical
features such as complex crystal 
structures \cite{li18}, rattling-like 
sublattice vibrations \cite{voneshen13, lin16}, lone-pair 
electrons \cite{Wang26, nielsen13}, ferroelectric
instabilities \cite{taneja25, sarkar20}, and
bonding hierarchies \cite{wu26}, etc.
Therefore, identifying and understanding
the microscopic mechanisms of thermal transport
in crystalline materials is crucial for 
designing materials with ultralow lattice
thermal conductivity. Phonons in crystals
carry heat through two microscopically distinct
mechanisms: particle-like propagation between
well-separated wave packets, and coherent, wave-like
tunnelling between near-degenerate modes whose
linewidths exceed their frequency
spacing \cite{Allen93}. The unified theory
of thermal transport formalizes this split into
a diagonal, Peierls-like contribution and an
off-diagonal coherence contribution \cite{Simoncelli19, Simoncelli22}.
The two channels respond to structural perturbations 
in opposite ways: phonon softening, mass disorder, or
structural complexity that suppresses $\kappa_p$ by 
shortening phonon lifetimes and lowering group 
velocities typically densify the phonon 
spectrum and narrow the interbranch spacing, thereby
enhancing $\kappa_c$ \cite{rodriguez26}. This 
intrinsic antagonism imposes a practical lower
bound on $\kappa_L$ in crystalline solids,
highlighting the need to decouple the two thermal transport
channels to further reduce lattice thermal conductivity.

Two recent studies show that this tradeoff is not
fundamental. Xiong et al. demonstrated that the
superionic crystals X$_6$Re$_6$S$_8$I$_8$ (X = Rb, Cs)
suppress both transport channels simultaneously 
through a bonding hierarchy \cite{xiong25}. A rigid
, covalently bonded [Re$_6$S$_8$I$_8$]$^{4-}$ 
covalent network is interpenetrated by weakly, ionically
bonded cations. The mechanism was further
generalized by Wu et al. into an explicit
structural descriptor - a phonon density of
states sparsity metric combined with a
mass contrast criterion \cite{wu26}. Together, these 
studies establish a soft ionically bound rattler
embedded in a stiff covalent framework as a general
route to simultaneous suppression of
particle- and wave-like transport channels.  

The AM$_3$Ch$_2$ (Where, A = alkali metal,
M = Cu, Ag, Au, and Ch = S, Se, Te) family of 
ternary chalcogenides provides a particularly 
compelling platform for exploring this 
concept \cite{Klepp96, rettie18, burschka177}. These
materials comprise strongly bonded [M$_3$Ch$_2$]$^-$ 
frameworks weakened by pronounced M-Ch antibonding 
character near the Fermi level, separated by a 
weakly bonded alkali metal sublattice. This
combination of a stiff, strongly bonded framework 
and weakly coupled, rattler-like sublattice is 
expected to promote distinct mechanisms for suppressing 
the particle-like and coherent component
of heat transport. Recent work on 
CsCu$_3$S$_2$ demonstrated a low lattice 
thermal conductivity, attributed primarily to
Cu-S $p-d$ antibonding hybridization and 
an avoided crossing between Cs-dominated
acoustic modes and low-lying optical
branches \cite{yue24}. Additionally, the sparse
high-frequency optical modes corresponding
to the strongly bonded lighter S atoms contribute
to the suppression of coherence's thermal 
conductivity. However, $\kappa_p$
remains substantial in CsCu$_3$S$_2$
with $\kappa_p$ of 1.8 and 1.37
W m$^{-1}$ K$^{-1}$ along in-plane and 
out-of-plane directions, respectively
at 300 K.
This contrast highlights a key
limitation of the bonding hierarchy picture: while
the weakly coupled sublattice can efficiently disrupt
particle-like transport, the strongly bonded covalent
framework may still promote substantial 
particle-like conduction, posing a key challenge
for simultaneous suppression of both thermal transport channels. 

This contrasting behavior reveals a fundamental
limitation of the bonding-hierarchy strategy
proposed for achieving ultralow thermal conductivity
by simultaneously suppressing both transport 
channels. While the weakly bonded sublattice 
suppresses particle-like transport through 
rattling-induced scattering and avoided-crossing-induced
flattening of phonon branches, the rigid 
covalent framework may still provide an efficient 
pathway for particle-like heat conduction, leaving
the particle-like transport channel 
only partially suppressed. Thus, simultaneous 
suppression of both transport
channels requires not only a weakly bonded 
rattler sublattice but also targeted engineering
of the strongly bonded covalent framework
itself to reduce its intrinsic capacity for
particle-like transport. 

The M-site cation therefore provides a natural 
target for further engineering. For 
instance, in CsCu$_3$S$_2$, the
S sublattice governs the spectral sparsity
that suppresses $\kappa_c$, while the 
weakly bonded Cs sublattice induces rattling
and avoided crossings that suppress $\kappa_p$. The
Cu–S bonds, which form the connective backbone 
of the covalent framework, remain 
comparatively untargeted and are responsible
for the residual particle-like
transport \cite{yue24}. To examine this possibility, we
replace Cu with its heavier group-11 congener
Ag. This is an isovalent substitution that preserves
the S-dominated high-frequency modes and
the Cs rattler sublattice, while increasing the
mass of the framework, forming M–Ch units. 
The increased mass is expected to soften the
acoustic and low-lying optical modes, reducing their
frequencies and group velocities and thereby
providing a chemically minimal route to further
suppress $\kappa_p$ without disrupting the mechanism
responsible for the low $\kappa_c$
in CsCu$_3$S$_2$. Similar mass-induced phonon 
softening and reduced group velocities have been
associated with suppressed thermal transport \cite{gusain26}.

In the present work, we systematically investigate
the structural stability, chemical bonding, anharmonic
lattice dynamics, and thermal transport properties of
CsAg$_3$S$_2$ in the trigonal $P\Bar{3}m1$
phase using first-principles simulations. We 
combine self-consistent phonon theory, including
quartic and cubic anharmonic renormalization, with
the unified theory of thermal transport. In
particular, the anharmonic phonon properties are
treated beyond the harmonic approximation by
incorporating both cubic and quartic 
anharmonicities into the phonon self-energy. The
unified thermal transport framework explicitly separates
the particle-like and coherent contributions to
lattice thermal conductivity. Together with the 
inclusion of higher-order phonon scattering 
and dual-channel phonon transport, we provide a 
comprehensive description of thermal 
transport in trigonal CsAg$_3$S$_2$. The 
formation energy and convex hull analysis 
confirm the structural stability. Our analysis 
further shows that the trigonal phase is dynamically
and mechanically stable and retains its structural
integrity up to 700 K. A pronounced bonding 
hierarchy was observed between the covalently 
bonded [Ag$_3$S$_2$]$^-$ framework and weakly 
coupled independent Cs$^+$ ions, resulting in an
ultralow lattice thermal conductivity 
of 0.222 W m$^{- 1}$ K$^{- 1}$ at 300 K. Importantly, both
particle-like and coherent transport channels are
strongly suppressed, while their relative contributions
change significantly with temperature. The
coherent contribution becomes dominant at 700 K, revealing
a crossover from particle-like to wave-like heat
transport. To complement the thermal 
transport analysis, we further investigate the
electrical transport properties within the 
momentum relaxation time approximation (MRTA). Combining
the ultralow lattice thermal conductivity with the 
resulting electrical transport coefficients, we obtain
a maximum ZT of 1.28 for $n$-type doping at 
700 K. Overall, our analysis demonstrates that 
targeted substitution on the covalent framework
within a bonding hierarchy structural motif can 
simultaneously suppress the two thermal 
transport channels, providing a new direction
for engineering ultralow thermal conductivity materials.

\section{Computational Methods}
\subsection{First-principles calculations}
We investigated the lattice dynamics, thermal, and
thermoelectric properties of CsAg$_3$S$_2$ using
first-principles calculations based on density
functional theory (DFT) as implemented in the
Vienna \textit{ab initio} simulation
package (VASP) \cite{Kresse96, KressePRB}. The 
projector augmented wave method with valence 
configurations $5s^{2}5p^{6}6s^{1}$ for 
Cs, $4d^{10}5s^{1}$ for Ag and $3s^{2}3p^{4}$ for
S was used to describe the interactions between
the ion cores and the valence electrons. 
The electronic wave functions were represented
using a plane-wave basis set with a 
kinetic energy cutoff of 500 eV. To describe
the electron-electron interactions, we utilized
the Perdew-Burke-Ernzerhof (PBE) exchange-correlation
functional \cite{Perdew96} within the generalized
gradient approximation (GGA). The conjugate gradient
algorithm was used to optimize the 
crystal structure by relaxing all degrees
of freedom with the energy and forces convergence
criteria of 10$^{-8}$ eV and 10$^{-7}$ eV/\AA, respectively.
The Brillouin zone was sampled using a k-mesh 
of $11\times11\times11$, while a denser mesh
of $21\times21\times21$ was utilized for the self-consistent
energy calculations. Spin-orbit coupling (SOC) was 
included in the electronic-structure calculations to account
for relativistic effects arising from the
presence of heavy elements. The crystal orbital
Hamilton population (COHP) analysis was performed
with the LOBSTER code \cite{Maintz16} to elucidate the 
nature of chemical bonds and orbital interactions.

\subsection{Interatomic force constants}
We utilized the ALAMODE package \cite{Tadano15, Tadano14} to 
efficiently extract the harmonic and anharmonic
interatomic force constants (IFCs). The 0 K
harmonic IFCs were obtained using the finite
displacement method. We utilized a $2 \times 2 \times 2$ 
supercell to create 12 displaced structures. In 
each structure, the atoms were 
displaced by 0.01~\AA. Subsequently, a
$5 \times 5 \times 5$ Monkhorst-Pack k-mesh
along with an energy cutoff of 500 eV was utilized
for the single-point DFT calculations. The
resulting displacement-force dataset was subsequently
used to extract the harmonic force constants
at 0 K. We obtained anharmonic IFCs by 
employing the compressive sensing lattice 
dynamics (CSLD) technique \cite{Zhou19, Zhou14}, which
can efficiently collect the physically important 
terms of the anharmonic IFCs by utilizing the
limited displacement-force dataset \cite{Tadano15}. We
conducted the \textit{ab initio} molecular 
dynamics simulations (AIMD) using a
$2 \times 2 \times 2$ supercell with a 
Monkhorst-Pack k-mesh of $2 \times 2 \times 2$ and
energy convergence criteria better than 10$^{-6}$ eV to
generate the displaced structures necessary to 
obtain the displacement and force dataset to 
compute the anharmonic force constants within the
framework of CSLD. The AIMD simulations
were carried out within the NVT ensemble using
a Nosé-Hoover thermostat at 300 K for 5000 steps
with a step size of 0.5 femto seconds (fs). After
disregarding 1000 initial steps, 80 snapshots were
sampled using the AIMD trajectory at equal 
intervals of 50 steps to generate the quasirandom 
configurations. Then a random displacement of 0.1~\AA{}
was further applied to all the atoms in the
sampled structures to minimize the
residual correlations between the structures.
Subsequently, high-precision static DFT 
calculations were performed to generate the training 
and validation data using the sampled 
configurations on a $5 \times 5 \times 5$ k-point
mesh with the energy convergence criteria of 10$^{-8}$ eV
and a plane-wave kinetic energy cutoff of 500 eV. Using
the displacement-force dataset and the 0 K harmonic 
force constants, the cubic and quartic IFCs were 
obtained using the least absolute shrinkage and selection
operator (LASSO) technique \cite{Nelson13, Candes08}. We
applied real-space cutoff radii of
7.94~\AA{} and 5.29~\AA{} for cubic and
quartic IFCs, respectively. Notably, we 
observed a low relative fitting error of 4.1 \% for 
anharmonic IFCs, which is in excellent 
agreement with the small values reported in the
literature \cite{Xia20, Zheng24, Tadano14, Tadano18}. This 
highlights the robustness and reliability of 
the present approach to accurately
model the anharmonic force constants.

\subsection{Anharmonic phonon renormalization}
The obtained 0 K harmonic and anharmonic IFCs are then
used to compute the temperature-renormalized phonon
dispersions within the framework of self-consistent 
phonon (SCP) theory as implemented in the 
ALAMODE package \cite{Errea11, Tadano18, Tadano15}. The
first-order SCP equation considering only the
quartic anharmonicity, i.e., the loop diagram, can be 
written in the following form \cite{Tadano22}
\begin{equation}
\begin{split}
    (\Omega^{S}_q)^{2} =  & (C^{\dagger}{_q} \Lambda^{(HA)}_{q} C_{q})_{\nu \nu} + 
    \frac{1}{N_{q^{'}}} \sum_{q^{'}} \Phi_{4}(q; -q; q^{'}; -q^{'}) \\
     &  \times  \frac{\hbar}{2\Omega^{S}_{q^{'}}}[1 + 2n(\Omega^{S}_{q^{'}})]
\end{split}
\end{equation}
Where $\Omega^{s}_q$ is the SCP renormalized phonon
frequency corresponding to wave vector q, $C_q$ is the 
unitary polarization matrix, $\nu$ represents the 
phonon branch index, $\Lambda^{(HA)}_{q}$ = diag$(\omega^{2}_{q1},...,\omega^{2}_{qs})$, $N_{q^{'}}$ is the total number 
of sampled phonon wave vectors in the first 
Brillouin zone, $\Phi_{4}(q; -q; q^{'}; -q^{'})$ is
the reciprocal representation of 4$^{th}$ order
IFCs and $n(\omega)$ is the Bose-Einstein distribution
function. The anharmonic polarization mixing was
explicitly considered in the calculations by including
the off-diagonal components associated with the loop self-energy. 
The SCP calculations were performed on 
a q-mesh of $2 \times 2 \times 2$, which corresponds
to the same supercell dimensions in real space. For 
the inner loop of the SCP equation, a denser
q-mesh of $8 \times 8 \times 8$ was utilized to 
obtain the converged anharmonic phonon frequencies.  \\
To accurately capture the effect of cubic
anharmonicity on the anharmonic phonon energies, we
further incorporated the bubble self-energy 
corrections to the anharmonic phonon energies (SCPB), which
can be evaluated using the following
self-consistent equation within the 
quasiparticle approximation \cite{Tadano22}
\begin{equation}\label{scpb}
     (\Omega^{B}_q)^{2} =  (\Omega^{S}_q)^{2} - 2 \Omega^{S}_{q}
     Re\sum^{B}_{q}[G^{S}, \phi_3](\omega)
\end{equation}
Where $\sum^{B}_{q}[G^{S}, \phi_3](\omega)$ is the phonon-frequency
dependent bubble self-energy defined as
\begin{equation}
\begin{split}
    \sum^{B}_{q} =  &\frac{1}{16} \sum_{q_{1}, q_{2}} 
    \frac{\hbar|\Phi(-q; q_{1}; q_{2})|^{2}}{\omega^{S}_{q} \omega^{S}_{q1} \omega^{S}_{q2}} \\
    & \times \Delta(-q + q{_1} + q_{2}) 
    f(1, 2, \omega + \iota \epsilon)
\end{split}
\end{equation}
Here, $\Phi_3$ represents the third-order IFCs
in reciprocal space, $\Delta(q)$ is a function
that becomes 1 when q is an integral multiple of the 
reciprocal lattice vector and 0 otherwise.
The $\omega$ dependent function $f(1, 2, \omega + \iota \epsilon)$
is given by 
\begin{equation}
\begin{split}
f(1, 2, \omega + \iota \epsilon) = \sum_{\sigma = -1, 1} & 
\sigma \Biggl[\frac{1 + n(\omega^{S}_{q1}) + 
n(\omega^{S}_{q2})}{\omega + \iota \epsilon + 
\sigma(\omega^{S}_{q1} + \omega^{S}_{q2})} \\
& - \frac{n(\omega^{S}_{q1}) -
n(\omega^{S}_{q2})}{\omega + \iota \epsilon +
\sigma(\omega^{S}_{q1} - \omega^{S}_{q2})} \Biggr]
\end{split}
\end{equation}
Where $n(\omega)$ is the Bose-Einstein distribution
function, $q_1$ and $q_2$ are the internal 
phonon wavevectors, and $\omega + \iota \epsilon$ represents
the analytic continuation of the Green's 
function to the real axis by a positive
infinitesimal $\epsilon$. Within the framework
of the quasiparticle (QP) approximation, several
treatments, including QP[0] ($\omega = 0$), 
QP[S] ($\omega = \Omega^{S}_q$), and the 
non-linear QPNL ($\omega = \Omega^{B}_q$) can be
implemented to solve Eq.~\eqref{scpb} \cite{Tadano22}. We
employed the QPNL scheme to obtain fully 
renormalized anharmonic phonon energies
accounting for both cubic and quartic
anharmonicities. A q-mesh of $2 \times 2 \times 2$ was
employed for the SCPB calculations. Since 
calculating the bubble self-energy is computationally 
expensive as compared to the SCP calculations, a
relatively coarse q-grid
of $4 \times 4 \times 4$ was used 
for the inner loop of the SCPB equation. Additionally,
the non-analytic part of the harmonic dynamical
matrix was included in all calculations. The
Born effective charges and dielectric 
tensor necessary to compute the non-analytic
correction were obtained using
density functional perturbation theory (DFPT) \cite{Baroni01}.

\subsection{Unified theory of thermal transport}
Utilizing the temperature-renormalized harmonic
and anharmonic IFCs, the lattice thermal 
conductivity was computed by
solving the linearized Boltzmann
transport equation (BTE) within the 
framework of the unified theory of 
thermal transport \cite{Simoncelli19, Simoncelli22}. Within
this formalism, the total lattice thermal
conductivity comprises the contribution
from $\kappa^{\alpha\beta}_p$ (capturing the 
particle-like phonon transport) 
and $\kappa^{\alpha\beta}_c$ (accounting for
the wave-like phonon tunneling) described by the diagonal 
and off-diagonal elements of the Wigner heat-flux 
operator, respectively \cite{Allen93}. The lattice 
thermal conductivity tensor under the 
single-mode relaxation-time approximation is given by
\begin{equation}
\begin{split}
    \kappa^{\alpha\beta}_{p/c} = & \frac{\hbar^{2}}{k_BT^{2}VN_{q}}\sum_{q}\sum_{j,j'}\frac{\Omega_{qj} +
    \Omega_{qj'}}{2}v^{\alpha}_{qjj'} \otimes v^{\beta}_{qj'j} \\
 \times & \frac{\Omega_{qj}n_{qj}(n_{qj} + 1)
 + \Omega_{qj'}n_{qj'}(n_{qj'} + 1)}{4(\Omega_{qj} 
 -\Omega_{qj'})^{2} + (\Gamma_{qj} + \Gamma_{qj'})^{2}}
 (\Gamma_{qj} + \Gamma_{qj'})
\end{split}
\label{kpac}
\end{equation}
Where $\hbar$, $k_B$, T, V, and N$_{q}$ represent 
the reduced Planck's constant, Boltzmann 
constant, absolute temperature, unit cell
volume, number of sampled $q$-points in the 
first Brillouin zone, and equilibrium Bose-Einstein
distribution function, respectively. $\Gamma_{qj}$
denotes the phonon scattering rate, which 
includes three-phonon ($\Gamma^{3ph}$), 
four-phonon ($\Gamma^{4ph}$) and 
isotope ($\Gamma^{isotope}$) scattering.  
$v_{qjj'}$ is the generalized group
velocity tensor including both intra- and
inter-branch contributions and is given by \cite{Allen93}
 \begin{equation}
    v_{qjj'} = \frac{\langle e_{qj}| 
    \frac{\partial D(q)}{\partial q} | e_{qj'} 
    \rangle}{2\sqrt{\Omega_{qj}\Omega_{qj'}}}
\end{equation}
with $e_{q}$ and D(q) representing the 
polarization vector and the dynamical matrix,
respectively. When only the diagonal elements of
the Wigner heat flux operator are 
considered, i.e., j = j' in the second summation, Eq.~\eqref{kpac}
reduces to the standard Peierls-Boltzmann contribution
to the lattice thermal transport given as
\begin{equation}
    \kappa^{\alpha\beta}_{p} = \frac{1}{VN_{q}}\sum_{qj}c_{qj}v^{\alpha}_{qj}v^{\beta}_{qj}\tau_{qj}
\end{equation}
Wherein, $c_{qj}$ and $v_{qj}$ are the modal heat
capacity and phonon group velocity, 
respectively. $\tau_{qj} = \frac{1}{2\Gamma_{qj}(\Omega_{qj})}$
represents the phonon relaxation time. Finally, the
total lattice thermal conductivity was 
obtained by summing the contributions 
from $\kappa^{\alpha\beta}_p$ and $\kappa^{\alpha\beta}_c$, 
i.e., $\kappa_L  = \kappa_p + \kappa_c$. In the 
present work, a q-mesh of $16 \times 16 \times 16$ along with
a scalebroad parameter of 0.5 was used for both 
3ph and 4ph scattering processes. Considering 
the memory-intensive nature of the computation
of 4ph scattering probabilities, the 
maximum likelihood approximation \cite{Guo24}  was used to
accelerate the convergence of the
thermal conductivity. It should be noted that 3ph
scattering processes are solved iteratively, while
the 4ph scattering processes are treated
under the relaxation time approximation. All thermal
transport calculations were carried out
utilizing the FourPhonon package along with 
its modified version to account for the coherent
conductivity \cite{Li14, Han22, Ji24}.

\subsection{Electrical transport properties}

We calculated the electrical transport parameters
by solving the electronic BTE within the 
momentum-relaxation-time approximation as
implemented in the AMSET 
package \cite{Ganose21}. The electron-phonon
interactions were modelled by explicitly 
considering the acoustic deformation potential (ADP), the
polar-optical phonon (POP) scattering, and the 
influence of ionized impurity (IMP) scattering. The
partial decay rate for each mechanism was 
obtained using Fermi's golden rule \cite{Dirac27}, and
the total scattering rate was calculated by
combining the following contributions
via Matthiessen's rule \cite{Matt64},
\begin{equation}
\frac{1}{\tau_\mathrm{e}} = \frac{1}{\tau_\mathrm{{ADP}}}
+ \frac{1}{\tau_\mathrm{{IMP}}} + \frac{1}{\tau_\mathrm{{POP}}}
\end{equation}
The input parameters, such as the electronic structure, 
dielectric tensor, elastic constants, deformation
potential, and polar optical phonon frequencies were obtained
from first-principles calculations.

\begin{figure}
\includegraphics[width=0.9\columnwidth]{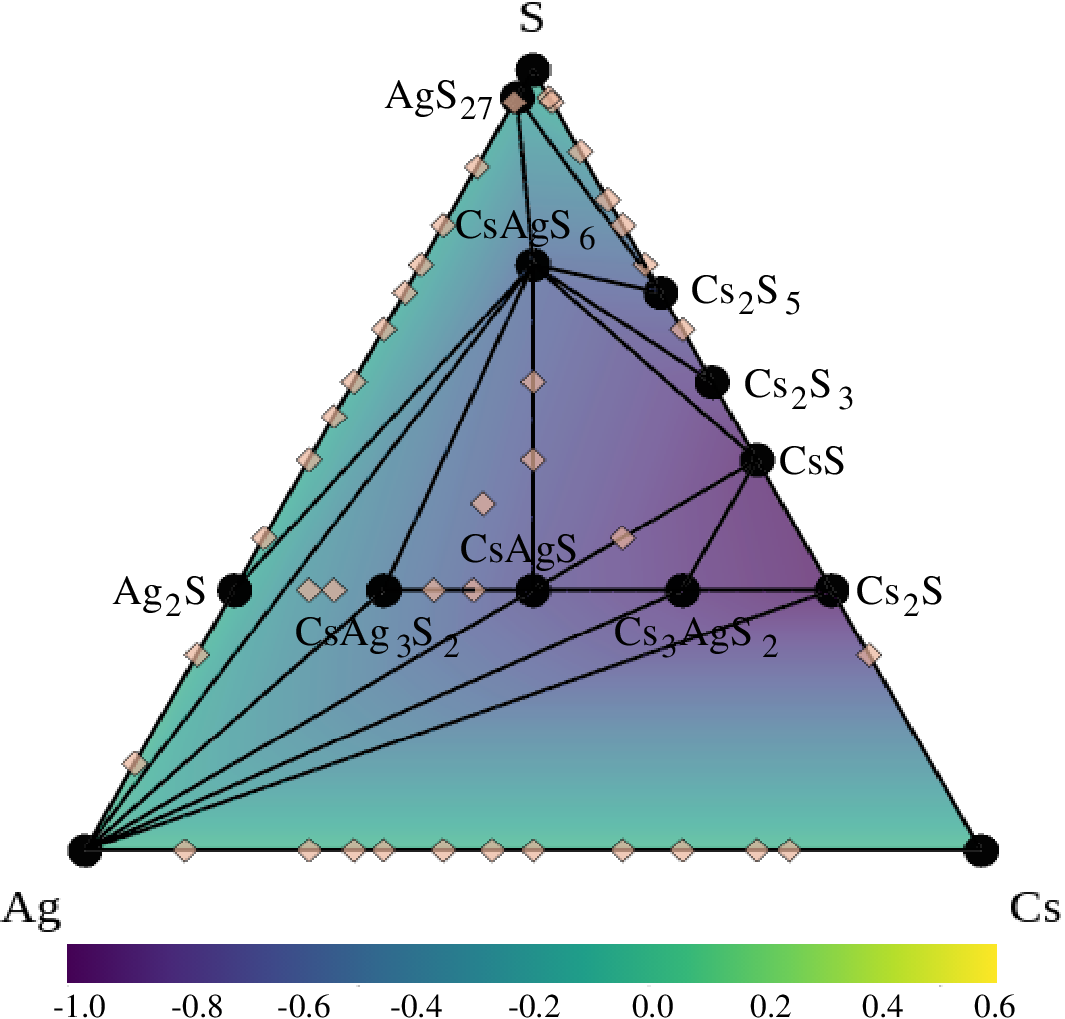}
\caption{Convex hull diagram for CsAg$_3$S$_2$ in trigonal phase. The color scale denotes the formation energy in eV/atom. Black circles indicate phases on the convex hull, while orange diamonds denote phases above the hull.}
\label{stability}
\end{figure}

\section{Results and Discussion}

\subsection{Structural Stability}
CsAg$_3$S$_2$ belongs to the  AM$_3$Ch$_2$ (Where, A
= alkali metal, M = Cu, Ag, Au and Ch = S, Se, Te)
family of ternary chalcogenides, whose members crystallize
in two closely related structure 
types: monoclinic $C2/m$ and trigonal $P\Bar{3}m1$, 
featuring corrugated [M$_3$Ch$_2$]$^-$ units 
separated by alkali cations \cite{burschka177, Klepp96}. 
CsAg$_3$S$_2$ was first synthesized in 1977 by 
the high-temperature melt reaction
method \cite{burschka177} and has since been
reported only in the monoclinic form. Despite
sharing the same bonding hierarchy, both
structure types differ in the spectral sparsity of
phonon modes they produce. The trigonal
cell contains single formula units, yielding a
phonon spectrum with well-separated optical
branches. In contrast, the large monoclinic cells
distribute twice as many branches within
a similar frequency range. The coherent phonon transport
becomes significant when the phonon 
linewidths become comparable to the interbranch spacing
between the phonon modes \cite{xie23}, and
therefore the monoclinic structure is prone to
enhanced wave-like thermal transport. The
trigonal structure already realized in Cu and Au 
members of the family \cite{Klepp96} therefore
provides a natural platform to test 
the hypothesis that targeted engineering of the
M-Ch framework can drive particle-like
transport into the ultralow regime while leaving
the coherent channel undisturbed.  

Before examining the lattice dynamics and
transport properties, it is essential to establish
the structural stability of
CsAg$_3$S$_2$ in the trigonal phase
and its energetic competitiveness with the
experimentally reported monoclinic phase. We
start the quest of stability by analyzing
the formation energy of both polymorphs. The formation
energies were computed using the expression 

\begin{equation}
    \Delta E_f = E_{CsAg_3S_2} - (E_{Cs} + 3E_{Ag} + 2E_{S})
\end{equation}
Where E$_{CsAg_3S_2}$ is the total energy per formula
unit, whereas $E_{Cs}$, $E_{Ag}$, and $E_{S}$ are
the energies of the respective elements per atom. Our
calculated formation energies for CsAg$_3$S$_2$ in
the trigonal and monoclinic phases
are -2.87 eV and -2.37 eV, respectively. Both
phases exhibit negative formation energies, indicating
stability against decomposition into 
constituent elements. The more negative
value of formation energy in the 
trigonal phase indicates that it is more
favorable energetically as compared to the 
experimentally observed monoclinic phase.

Having established the energetic favorability 
of both polymorphs, we next assess their 
thermodynamic stability against competing phases 
in the Cs-Ag-S chemical space through 
convex hull analysis. The convex hull provides
a rigorous measure of the stability of the
material with respect to all possible combinations
of competing phases \cite{Bartel22}. Phases
lying on the convex hull are thermodynamically
stable, whereas materials lying above the convex hull
are prone to decomposition into the competing
phases and therefore thermodynamically unstable. We
constructed the convex hull using all available
binary and ternary phases from the OQMD 
database \cite{saal13, kirklin15}, including
experimentally reported compounds and 
theoretically predicted structures based on common
structure prototypes. A total of 48 competing
phases were optimized using identical computational
parameters employed for CsAg$_3$S$_2$, followed
by static calculations to obtain their respective
formation energies and ensure computational
consistency. As shown in Fig.S3 \cite{suppl}, the
trigonal phase lies on the convex hull, indicating
its thermodynamic stability at 0 K. In contrast, the
experimentally reported monoclinic phase
lies 0.057 eV/atom above the hull. This relatively
small hull distance is consistent with the experimental 
realization of the monoclinic phase \cite{ma17}. The
convex hull analysis therefore establishes
the thermodynamic stability of the trigonal phase
against decomposition into competing phases at 0 K.   


\begin{figure}
\includegraphics[width=\columnwidth]{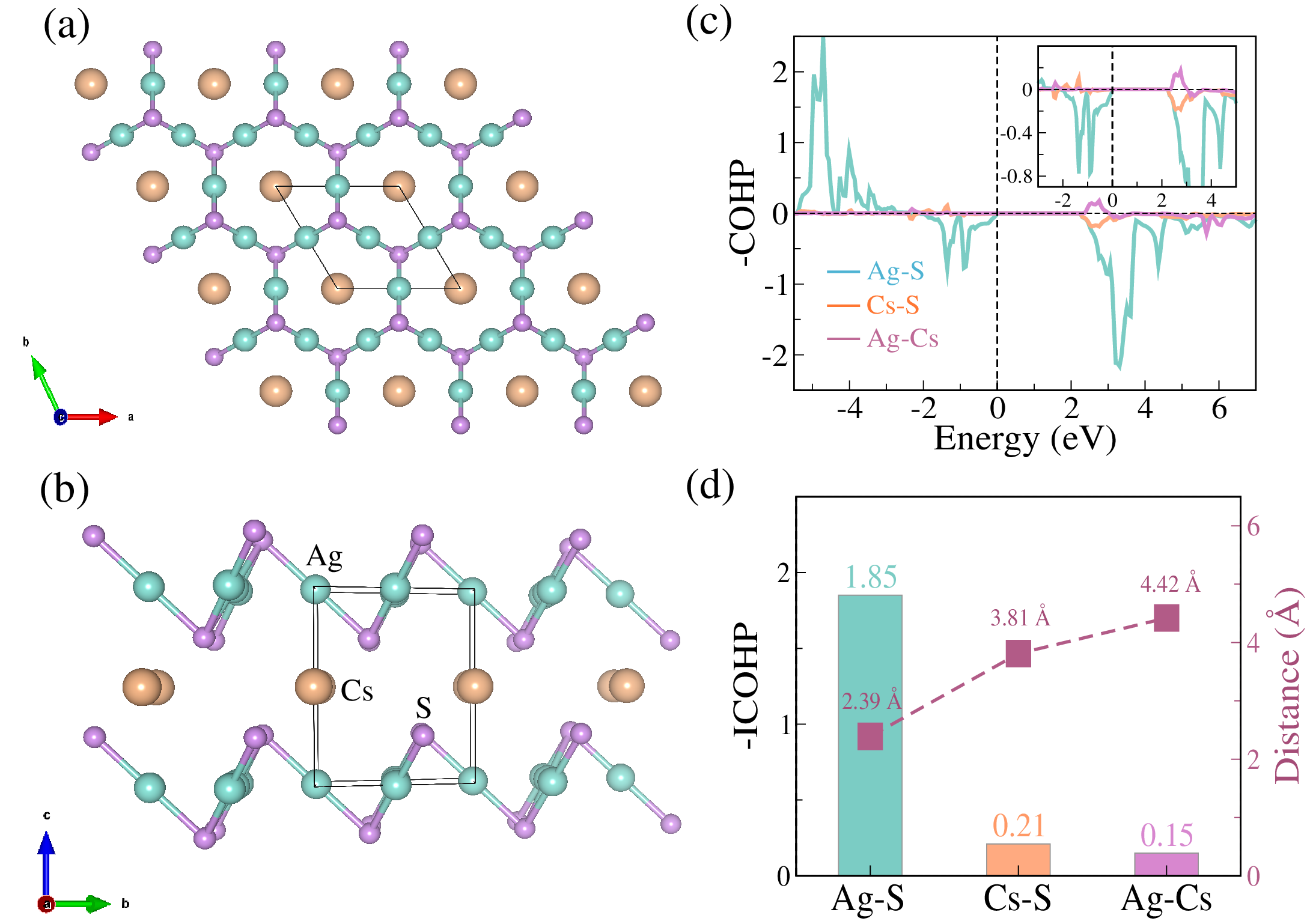}
\caption{(a) Top view of CsAg$_3$S$_2$ crystal structure in the
trigonal phase. (b) Corrugated Ag-S layers
running parallel to the (001) direction. (c) Calculated crystal orbital
Hamilton population (ICOHP) for CsAg$_3$S$_2$. The inset shows
the enlarged view near the Fermi level featuring the
Ag-S antibonding interactions. (d) Integrated COHP and 
bond distance analysis for CsAg$_3$S$_2$.}
\label{crystal}
\end{figure}

\begin{figure*}[t]
\centering
\includegraphics[width=\textwidth]{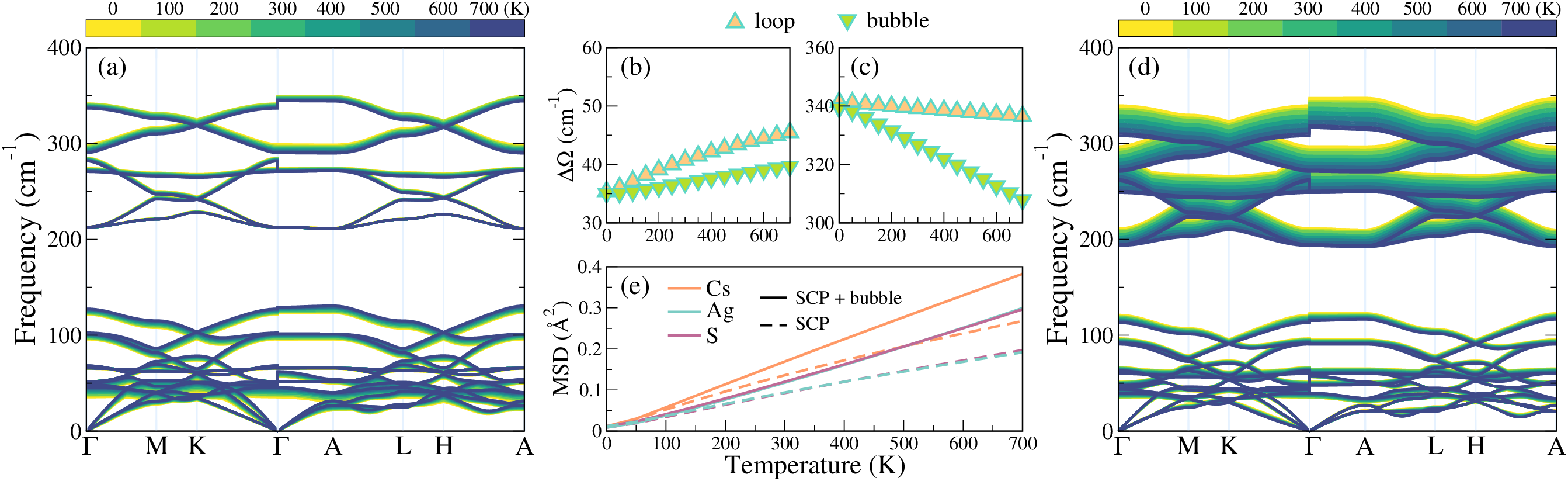}
\caption{(a) Phonon dispersions for CsAg$_3$S$_2$ including
quartic anharmonicity (loop diagram). (b) Temperature
dependence of the lowest and (c) highest
optical modes using loop and bubble 
diagrams. (d) Phonon dispersion for
CsAg$_3$S$_2$ including both cubic (bubble 
diagram) and quartic anharmonicity (loop diagram). 
(e) Calculated mean-square displacements for 
Cs, Ag, and S atoms within the SCP and SCPB 
frameworks, respectively. Note that Ag and S values overlap.}
\label{phonons}
\end{figure*}

Establishing the thermodynamic stability, we 
conducted the AIMD simulations at 300 K, 500 K,
and 700 K to verify the structural stability at
higher temperatures. As shown in Fig. S2 (b) \cite{suppl} 
the free energy fluctuated only slightly during
the 20 ps AIMD run, and no evidence of structural
distortion or phase transitions was observed
. This suggests that CsAg$_3$S$_2$ can withstand 
higher temperatures relevant to the practical
operating conditions.

Having confirmed the thermodynamic and thermal
stability of trigonal CsAg$_3$S$_2$, it is equally
important to examine its mechanical stability,
which is necessary for assessing a material's ability
to withstand external stresses and
mechanical load without undergoing permanent
structural deformation or collapse. Our computed elastic
constants, listed in Table X, obey the 
Born–Huang elastic stability criteria given by 
\begin{subequations}
\begin{equation}
C_{11} > |C_{12}|, C_{44} > 0  
\end{equation}
\begin{equation}
2C^{2}_{13} < C_{33}(C_{11} + C_{12})
\end{equation}
\begin{equation}
2C^{2}_{14} < C_{44}(C_{11} - C_{12})
\equiv C_{44}C_{66}
\end{equation}
\end{subequations}
The above findings underscore the mechanical
robustness of CsAg$_3$S$_2$. Complementing 
this, the absence of imaginary frequencies
in the phonon dispersions [see Fig.~\ref{phonons} (a) and
(e)] establishes the dynamical stability, thereby ruling
out any possibility of soft-mode-driven
structural phase transition. Collectively, the
favorable formation energy, thermodynamic
stability against competing phases as evidenced
by the convex hull analysis, finite-temperature
structural integrity under AIMD simulations, and
the fulfillment of mechanical and dynamical
stability criteria provide compelling evidence for
the intrinsic stability of the trigonal phase 
of CsAg$_3$S$_2$. Having firmly established the 
stability of CsAg$_3$S$_2$ in the trigonal
symmetry, we now turn our attention to its crystal
structure and bonding characteristics.

\begin{figure}
\includegraphics[width=0.95\columnwidth]{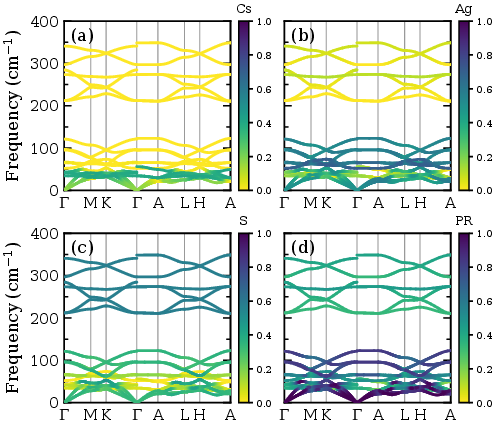}
\caption{Atomic participation ratio for (a) Cs, (b) Ag, and
(c) S atoms at 300 K. (d) Total participation
ratio for CsAg$_3$S$_2$ at 300 K.}
\label{apr}
\end{figure}

\begin{figure*}
\includegraphics[width=0.95\textwidth]{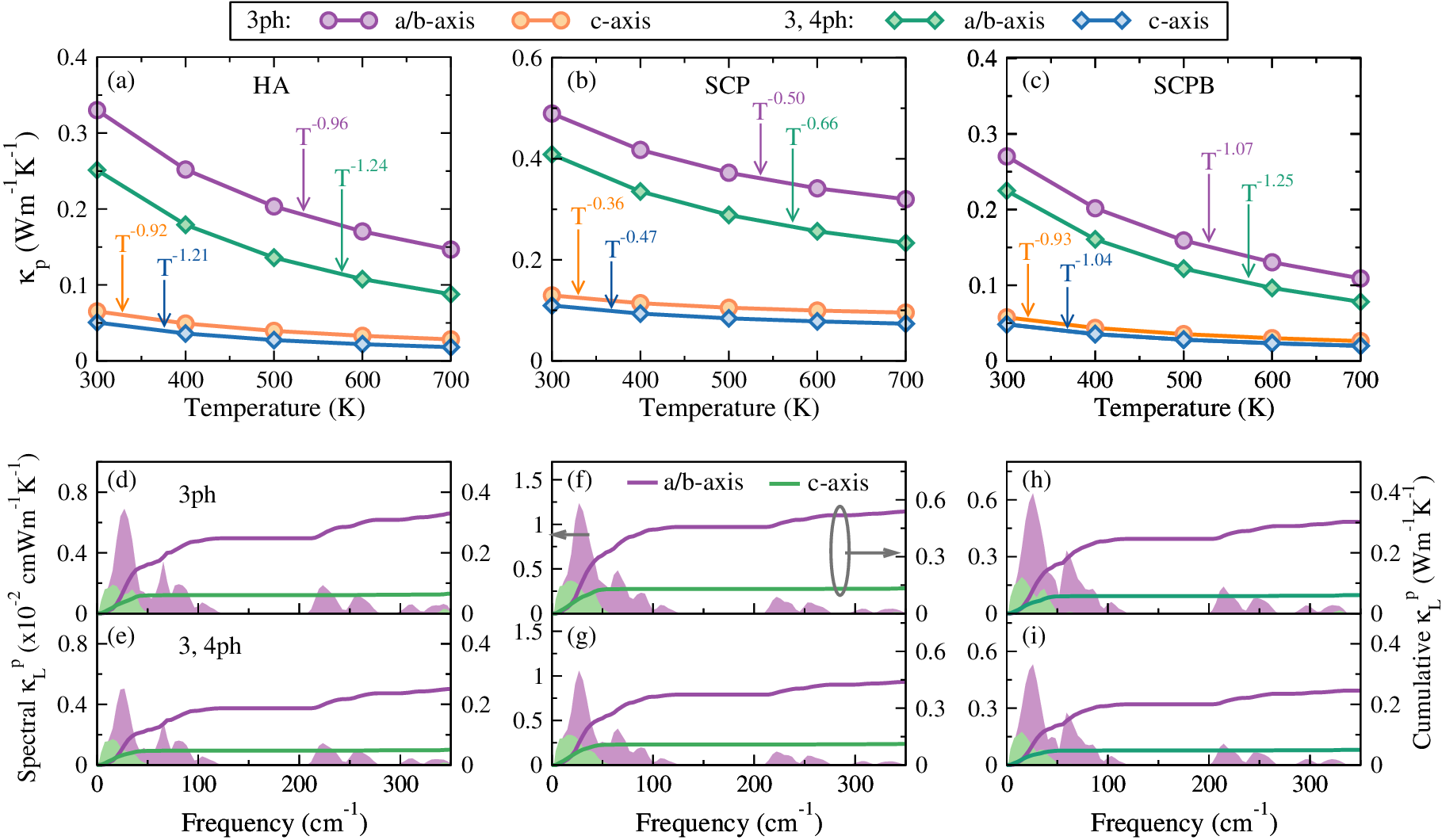}
\caption{Temperature dependence of particle-like
thermal conductivity ($\kappa_p$) for CsAg$_3$S$_2$ along 
in-plane and out-of-plane
directions within (a) HA, (b) SCP, and (c) SCPB 
frameworks considering 3ph (circles)
and 3,4ph (diamonds) processes.
The downward arrows represent the 
fitted temperature dependence
$\kappa_p \propto T^{-\alpha}$. 
(d) Spectral (shaded, left axis)
and cumulative (line, right axis) $\kappa_p$
at 300 K along in-plane and out-of-plane
directions considering 3ph processes 
within the HA framework. (e) Same as (d) but
considering 3,4ph processes. (f) Same as (d)
but within the SCP framework. (g) Same as
(e) but within the SCP framework. (h) Same
as (d) but within the SCPB framework. (i) Same
as (e) but within the SCPB framework.} 
\label{kpa}
\end{figure*}

\begin{figure*}
\includegraphics[width=0.95\textwidth]{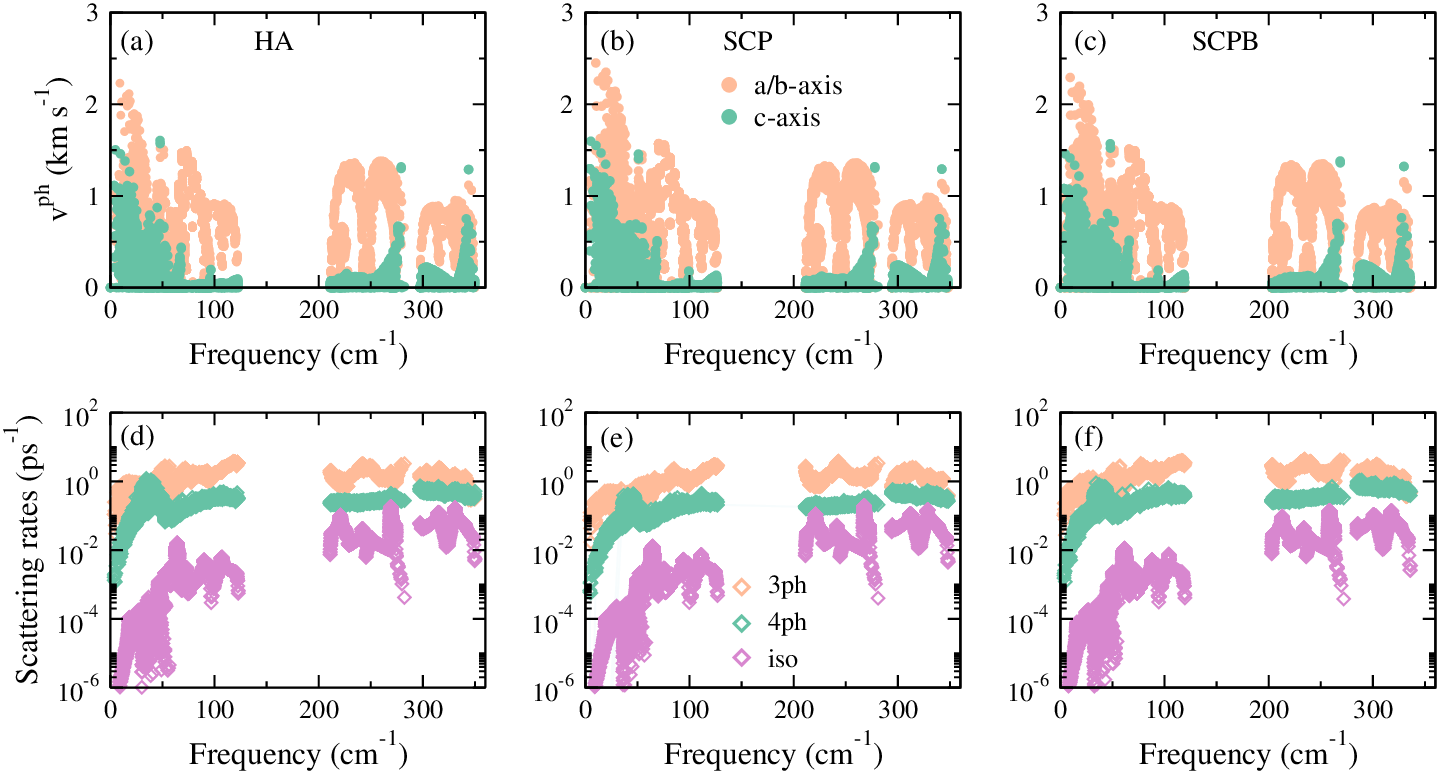}
\caption{(a) Phonon group velocities as a function of
frequency at 300 K along in-plane
and out-of-plane directions within the HA framework. 
(b) Same as (a) but within the SCP
framework. (c) Same as (a) but 
within the SCPB framework. (d) Three-phonon (3ph),
four-phonon (4ph), and isotope (iso) scattering rates
as a function of frequency at 300 K within
the HA framework. (e) Same as (d) but within the SCP 
framework. (f) Same as (d) but within the SCPB framework.}
\label{scattering}
\end{figure*}

\subsection{Crystal Structure and Bonding}
The crystal structure of trigonal CsAg$_3$S$_2$ is
shown in Fig.~\ref{crystal} (a) and (b). 
CsAg$_3$S$_2$ crystallizes in 
the $P\Bar{3}m1$ space group (No. 164) with 6 atoms
in the primitive cell, where the Cs, Ag, and S
occupy the 1b, 3e, and 2d Wyckoff 
positions, respectively. The optimized lattice
parameters are listed in 
Table. S1 \cite{suppl}. The structure
is isostructural with CsCu$_3$S$_2$ \cite{sato92} and
consists of corrugated [Ag$_3$S$_2$]$^-$ layers
parallel to the (001) plane, separated
by Cs$^+$ ions. Within this structure, the Ag
atoms are in strictly linear coordination, while
S is triply coordinated within the Ag-S
framework. In contrast, Cs is surrounded by S atoms
forming a distorted octahedral environment [Fig. S4 (a) \cite{suppl}], with 
a Cs-S distance of 3.81 \AA and S-Cs-S angles
of 76.6$^\circ$ and 103.4$^\circ$. The distorted
Cs environment, together with the large mass of Cs
and its weak interaction with the Ag-S framework, is
expected to produce soft, low-frequency
Cs-dominated modes and strong lattice anharmonicity. 

To further understand the bonding characteristics, we 
performed crystal orbital Hamilton population
analysis as shown in Fig.~\ref{crystal} (c) and (d).
The Ag-S interactions
exhibit pronounced bonding states between
-4 eV to -6 eV, with an integrated COHP (ICOHP)
value of -1.85 eV, confirming the strong covalent
character of the Ag-S framework. In comparison, the
Cs-S and Cs-Ag interactions are considerably
weaker, with ICOHP values of -0.21 eV and -0.15 eV, respectively.
For the isostructural CsCu$_3$S$_2$, when computed within
identical input parameters, the corresponding Cu-S
and Cs-S values are -2.35 eV and -0.75 eV, 
respectively [see Fig. S5 in SI \cite{suppl}]. This indicates
that Cu to Ag substitution not only weakens the
covalent framework but also diminishes the 
Cs-S interaction strength arising from 
the enlargement of the CsS$_6$ cavity and increased
Cs-S bond lengths. Notably, antibonding 
Ag-S states occur near the Fermi level
due to Ag-$d$ and S-$p$ hybridization. 
Although these states represent only 
a small fraction of the occupied manifold, their 
presence provides an additional source of 
anharmonicity while the overall Ag-S framework
remains strongly bonded \cite{yuan23}.  

This bonding picture is further supported by Bader
charge analysis. Cs, Ag, and S carry charges
of +0.84 $e$, +0.19 $e$, and -0.70 $e$, 
respectively, indicating substantial charge
transfer from Cs and appreciable charge sharing
between Ag and S. Thus, the lattice exhibits a clear
bonding hierarchy consisting of a strongly 
bonded covalent Ag-S framework with a weakly coupled
Cs sublattice. The large contrast in bond 
strength and atomic mass is expected to produce distinct
vibrational characteristics of Cs and Ag-S
sublattices, providing the structural basis for the
anomalous lattice dynamics and ultralow thermal
conductivity as discussed in subsequent sections.

\subsection{Anharmonic Lattice Dynamics}

The temperature-renormalized phonon
dispersions of CsAg$_3$S$_2$ are presented
in Fig.~\ref{phonons}. Upon inclusion of the quartic
anharmonicity through the loop diagram, the 
phonon spectrum exhibits a strong mode-dependent
temperature response. The acoustic and low-frequency 
optical branches below 130 cm$^{-1}$ progressively
harden with increasing temperature, whereas the
intermediate frequency modes between 212 and
280 cm$^{-1}$ remain nearly unchanged. In 
contrast, the high-frequency optical modes soften
at elevated temperatures. This contrasting 
behavior reflects the distinct contribution of quartic
anharmonicity to different regions of the
phonon spectrum. The loop-only contribution to
phonon self-energy is known to overestimate
the phonon frequencies \cite{zheng22, Tadano22}. We 
therefore include the cubic anharmonicity through the
bubble diagram to obtain fully renormalized phonon
frequencies \cite{Tadano22}. As shown
in Fig.~\ref{phonons} (e), the inclusion of
the bubble contribution substantially 
modifies the temperature evolution of the phonon
spectrum. The combined effect of the
loop and bubble self-energies determines the
net temperature renormalization. The 
high-frequency optical modes undergo a 
pronounced softening because the cubic and
quartic contributions act cooperatively. The 
low-frequency optical modes also demonstrate an 
overall softening behavior; the temperature 
dependence is, however, weak due to the competing 
effects of quartic and cubic anharmonicities. Similar
mode-dependent temperature renormalization was
also observed in CsCu$_3$S$_2$ \cite{yue24}, 
CsCu$_4$Se$_3$ \cite{yue25}, and Cs$_3$Cu$_2$I$_5$ \cite{tao25}.

The contrasting roles of cubic and quartic 
anharmonicity are further illustrated by tracing
the lowest and highest frequency optical 
modes, M$_1$ and M$_2$, respectively
[See Fig.~\ref{phonons} (b), (c)]. The M$_1$ mode
hardens with temperature under both quartic
and cubic anharmonicities, although the bubble
contribution reduces its temperature dependence. In 
contrast, M$_2$ softens under both treatments, with
the bubble contribution substantially
enhancing its temperature dependence. These results
indicate that the mid- and high-frequency
modes are strongly influenced by cubic 
anharmonicity, whereas the low-frequency optical
modes exhibit a significant dependence on 
quartic anharmonicity. Compared to the isostructural 
CsCu$_3$S$_2$, which exhibits the softening of 
high-frequency optical modes along with 
the hardening of the acoustic and low-frequency
optical modes \cite{yue24}, CsAg$_3$S$_2$ exhibits
contrasting temperature dependence.  Both
the high- and low-frequency optical modes show
appreciable softening. Notably, the acoustic
modes show a decrease in frequency as the temperature
increases, indicating that the Ag-S framework
can effectively suppress the group velocities of
these modes and is therefore expected to
lower the particle-like thermal conductivity.

\begin{figure*}
\includegraphics[width=0.95\textwidth]{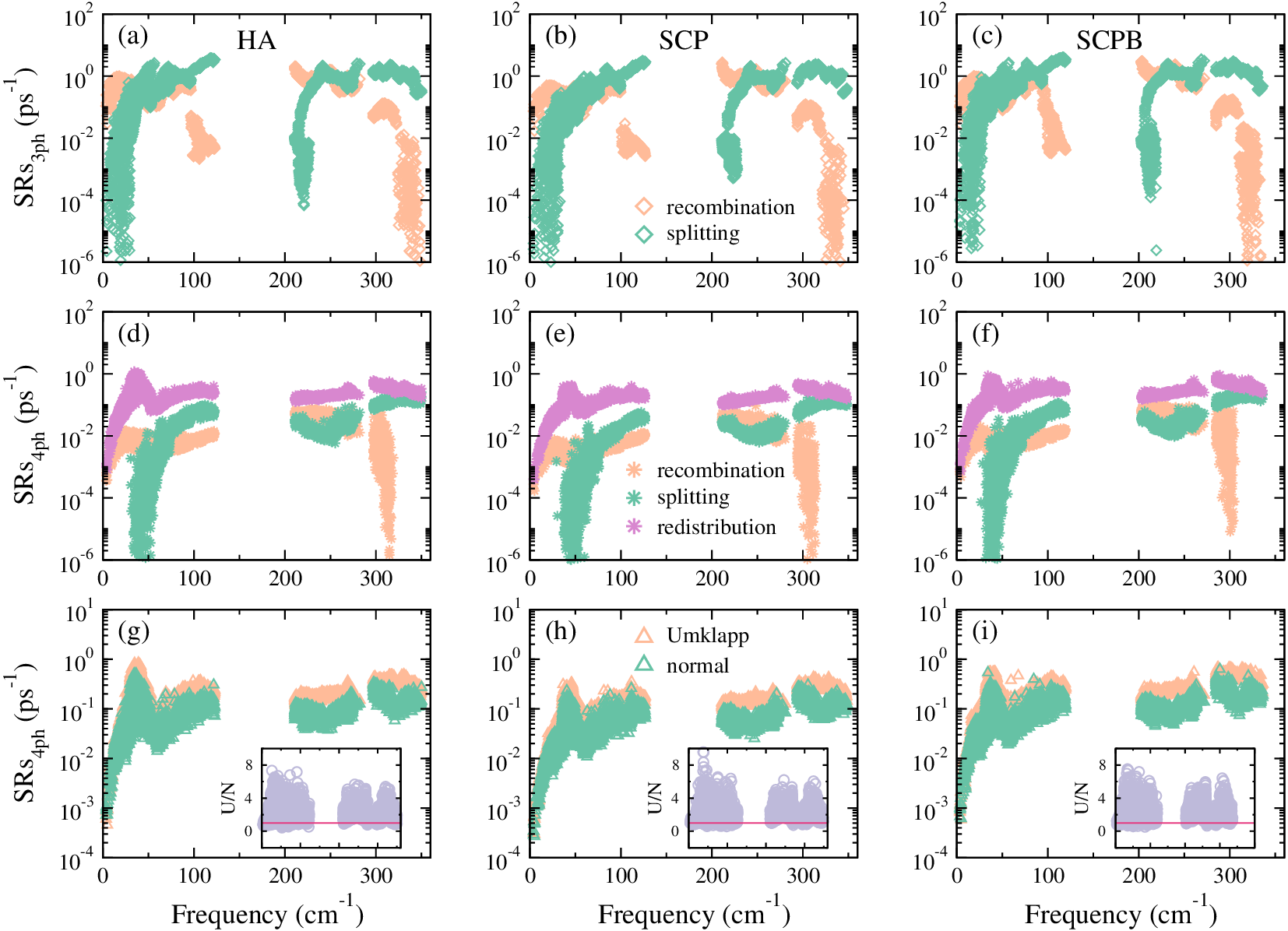}
\caption{(a) Process-resolved 3ph scattering 
processes as a function of frequency at 300 K
within the HA framework. (b) Same as (a) but
within the SCP framework. (c) Same as (a) but
within the SCPB framework.
(d) Process-resolved 4ph scattering processes
as a function of frequency at 300 K within
the HA framework. (e) Same as (d) but within
the SCP framework. (f) Same as (d) but within
the SCPB framework. (g) Umklapp (U)
and normal (N) contributions to the 4ph
processes as a function of frequency at 
300 K within the HA framework. The inset 
shows the ratio of Umklapp to normal scattering
processes (U/N). (h) Same as (g) but within
the SCP framework. (i) Same as (g) but within
the SCPB framework.}
\label{subprocessscattering}
\end{figure*}

The atom-resolved phonon density of states provides
further insights into the microscopic origin 
of the unusual lattice dynamics in CsAg$_3$S$_2$,
as shown in Fig. S6 \cite{suppl}. The distinct mass and bonding
hierarchy of Cs, Ag, and S results in a clear
separation of their vibrational contributions
across the phonon spectrum. The acoustic branches
are dominated by Cs vibrations, while 
the low-lying optical modes in the frequency
range 50-150 cm$^{-1}$ arise
predominantly from Ag atoms. In contrast, the 
high-frequency optical manifold above the 
phonon gap, extending from 210-350 cm$^{-1}$, is
dominated by the lighter S atoms. Additionally, an
avoided crossing between the acoustic and
low-lying optical branches is observed along the
$\Gamma$-A direction. Such hybridization is 
characteristic of rattling-like motion of the 
loosely bound Cs atoms, which is further supported
by the larger mean square displacement (MSD) values
of Cs atoms compared to Ag and S atoms
[see Fig.~\ref{phonons} (e)] and weak Cs-S 
interactions. This hybridization leads
to the branch flattening of the interacting modes
and consequently reduces phonon group velocities \cite{lin16}. 

To elucidate the microscopic nature of the 
vibrational modes, we computed the atomic
participation ratio (APR) and participation 
ratio (PR) \cite{hafner93}, defined as 
\begin{subequations}
    \begin{equation}
        APR_{qj,\kappa} = \frac{|e(\kappa;qj)|^2}{M_\kappa} 
        \Bigg/ \left( N_\kappa \sum_{\kappa=1}^{N_\kappa}
        \frac{|e(\kappa;qj)|^4}{M_\kappa^2} \right)^{1/2},
    \end{equation}
    \begin{equation}
        PR_{qj} = \left( \sum_{\kappa}^{N_\kappa}
        \frac{|e(\kappa;qj)|^2}{M_\kappa}
        \right)^2 \Bigg/ N_\kappa \sum_{\kappa=1}^{N_\kappa}
        \frac{|e(\kappa;qj)|^4}{M_\kappa^2},
    \end{equation}
\end{subequations}

where $\kappa$ is the atom index, $N_{\kappa}$ denotes
the number of atoms, $e(\kappa; qj)$ represents 
the phonon eigenvector with wave number $q$ and 
mode index $j$, and $M_\kappa$ is the atomic
mass. The APR quantifies the contribution of individual
atoms to a particular phonon mode, whereas PR
can be used to distinguish extended propagative 
modes from localized vibrations.

As shown in Fig.~\ref{apr} (a)-(c), the vibrational 
modes below 50 cm$^{-1}$ are predominantly 
associated with Cs, whereas Ag dominates the
low-lying optical modes spanning from 50-150
cm$^{-1}$. Above the phonon gap, the high-frequency
optical modes are dominated by S, consistent
with its low mass and strong incorporation into the 
covalent framework. Notably, the Cs contribution
rapidly vanishes above 50 cm$^{-1}$, indicating 
its confined motion to a narrow frequency window rather
than being distributed throughout the 
spectrum. Additionally, the PR shown in Fig.~\ref{apr} (d) further
resolves the character of the low-frequency manifold. The
acoustic modes exhibit PR values close to unity, characteristic 
of extended propagative vibration, whereas the PR
decreases sharply above 25 cm$^{-1}$ with pronounced
localization in the 25-50 cm$^{-1}$ frequency 
window. Thus, the Cs sublattice contributes to both the
acoustic modes and the localized vibrations
immediately above them. The avoided crossing between 
the Cs-derived modes hybridizes the localized optical 
character with the acoustic branches, leading to 
branch flattening, reduced group velocities, and enhanced 
anharmonic scattering. A similar mechanism of mode 
localization and suppression of particle-like thermal 
transport has been reported in filled
skutterudites \cite{pang22}, type-I 
clathrates \cite{pailhes14}, and related 
chalcogenides such as CsCu$_3$S$_2$ \cite{yue24} 
and CsCu$_4$Se$_3$ \cite{yue25}.

Taken together, the lattice dynamics of 
trigonal CsAg$_3$S$_2$ reveal complementary 
mechanisms for suppressing particle-like thermal
transport: Cs-derived localized modes enhance 
anharmonic scattering and shorten phonon lifetimes, while
the heavy Ag atoms in the more weakly bonded
Ag-S framework reduce
the group velocity of heat-carrying phonon modes. 
Compared with the isostructural CsCu$_3$S$_2$, the
Cu to Ag substitution enhances both mechanisms by
weakening the covalent framework and loosening
the Cs rattlers. The combined influence of 
shortened phonon lifetimes and diminished group
velocities leads to a substantial reduction in 
particle-like thermal
transport, as quantified below.

\subsection{Lattice Thermal Conductivity}
To elucidate the interplay between anharmonic
phonon renormalization and multiphonon scattering in
governing lattice thermal transport in CsAg$_3$S$_2$, we
calculated the lattice thermal conductivity using 
temperature-renormalized phonon frequencies obtained within
the HA, SCP, and SCPB frameworks together with 
three-phonon (3ph) and combined three- and
four-phonon (3,4ph) scattering processes. Consequently, we
constructed six theoretical models, namely HA+3ph,
HA+3,4ph, SCP+3ph, SCP+3,4ph, SCPB+3ph, and SCPB+3,4ph. As
shown in Fig.~\ref{kpa} (a-c), all six approaches 
predict pronounced anisotropy in $\kappa_p$, with
the in-plane component substantially exceeding
the out-of-plane component. 
Within the SCPB+3,4ph framework, 
$\kappa_p$ reaches 0.22 and
0.05~ W m$^{-1}$ K$^{-1}$ along the $a/b$-
and $c$-axis at 300~K,
as compared to 1.80 and
1.37 W m$^{-1}$ K$^{-1}$ for the isostructural
CsCu$_3$S$_2$ \cite{yue24}.
The in-plane reduction reflects
the heavier and more weakly bonded 
Ag--S framework, whereas the much 
stronger suppression along $c$-axis is
consistent with the considerably weaker
Cs--S interaction, since heat transport
across the layers is mediated
by the Cs--S contacts.

Fig. ~\ref{kpa} (a)-(c) further illustrates
the systematic temperature dependence of $\kappa_p$ as 
predicted by different theoretical models. Within
the 3ph approximation, the SCP+3ph model yields
the highest $\kappa_p$ (0.49 and 0.13 W/mK along 
the a/b- and c-axis, respectively, at 300 K), followed
by HA+3ph (0.33 and 0.07 W/mK) and 
SCPB+3ph (0.27 and 0.06). 
The enhancement
within SCP+3ph framework arises from
the quartic-induced hardening
of the acoustic and
low-frequency optical modes,
which increases the group
velocities, whereas the bubble
self-energy in SCPB+3ph framework
softens these modes and
markedly suppresses $\kappa_p$.

The inclusion of 4ph scattering
systematically reduces $\kappa_p$ 
in both directions throughout the
investigated temperature range. The 
reduction becomes progressively larger with 
increasing temperature, indicating 
the growing importance of higher-order 
anharmonic scattering at elevated
temperatures. The in-plane
$\kappa_p$ decreases by 24--40\%, 16.5--27\%, 
and 17--29\% within HA, SCP, and SCPB, respectively,
with corresponding out-of-plane 
reductions of 23--36\%, 15--23\%, and 16--24\%.
These values are comparable to those reported for
Cs$_2$AgBiBr$_6$ (34.6\% and 51.8\% at 300 and 700~K)
\cite{Zheng24} and CsCu$_4$Se$_3$ 
($\sim$46\% in-plane and $\sim$38\% cross-plane at
300~K) \cite{yue25}. Notably, the influence of 
4ph scattering is consistently stronger
within the HA approximation, whereas the results are
comparatively smaller for the SCP and 
SCPB frameworks, suggesting  that
temperature-induced renormalization 
partially compensates for additional thermal resistance 
introduced by the 4ph scattering channels
by modifying the phonon spectrum and the 
scattering phase space. Nevertheless, the inclusion of 
4ph scattering does not alter the pronounced 
anisotropy of $\kappa_p$, but primarily lowers 
its magnitude by introducing additional resistive
 phonon scattering processes that shorten
 the phonon lifetimes.  

To quantify the temperature dependence of
the lattice thermal conductivity, we fitted $\kappa_p$ 
to the power law expressed as
$\kappa_p \propto T^{-\alpha}$, where $\alpha$ can be 
used as an effective descriptor of the temperature
evolution of underlying phonon scattering
mechanism. Within the HA+3ph framework, $\kappa_p$ exhibits
a temperature dependence of $T^{-0.96}$ $(T^{-0.92})$ along 
a/b-axis (c-axis), close to the classical $T^{-1}$ dependence
of Umklapp-limited heat transport. In contrast, SCP+3ph
yields a much weaker temperature dependence
of $T^{-0.50}$ $(T^{-0.36})$, significantly diverging 
from the $T^{-1}$ scaling, reflecting the partial 
compensation of the temperature-induced
increase in scattering by quartic-anharmonic phonon
renormalization. 
Including the bubble correction
restores the stronger temperature 
dependence of $T^{-1.07}$ ($T^{-0.93}$) within the
SCPB+3,4ph approach, consistent with the
phonon softening induced by cubic anharmonicity.
With 4ph scattering,
the exponents evolve to $T^{-1.24}$ ($T^{-1.21}$),
$T^{-0.66}$ ($T^{-0.47}$), and $T^{-1.25}$ ($T^{-1.04}$)
within HA+3,4ph, SCP+3,4ph, and SCPB+3,4ph,
respectively. The obtained in-plane SCPB+3,4ph
exponent is comparable to
that of CsCu$_3$S$_2$
($T^{-1.05}$) \cite{yue24},
indicating that the Cu to Ag
substitution mainly lowers
the magnitude of $\kappa_p$
rather than changing its
temperature scaling. 
Such deviations from classical $T^{-1}$ behavior are
also reported in other strongly anharmonic 
materials \cite{Xiao25, song25, Virakante26, Zheng24}.

\begin{figure*}
\includegraphics[width=0.9\textwidth]{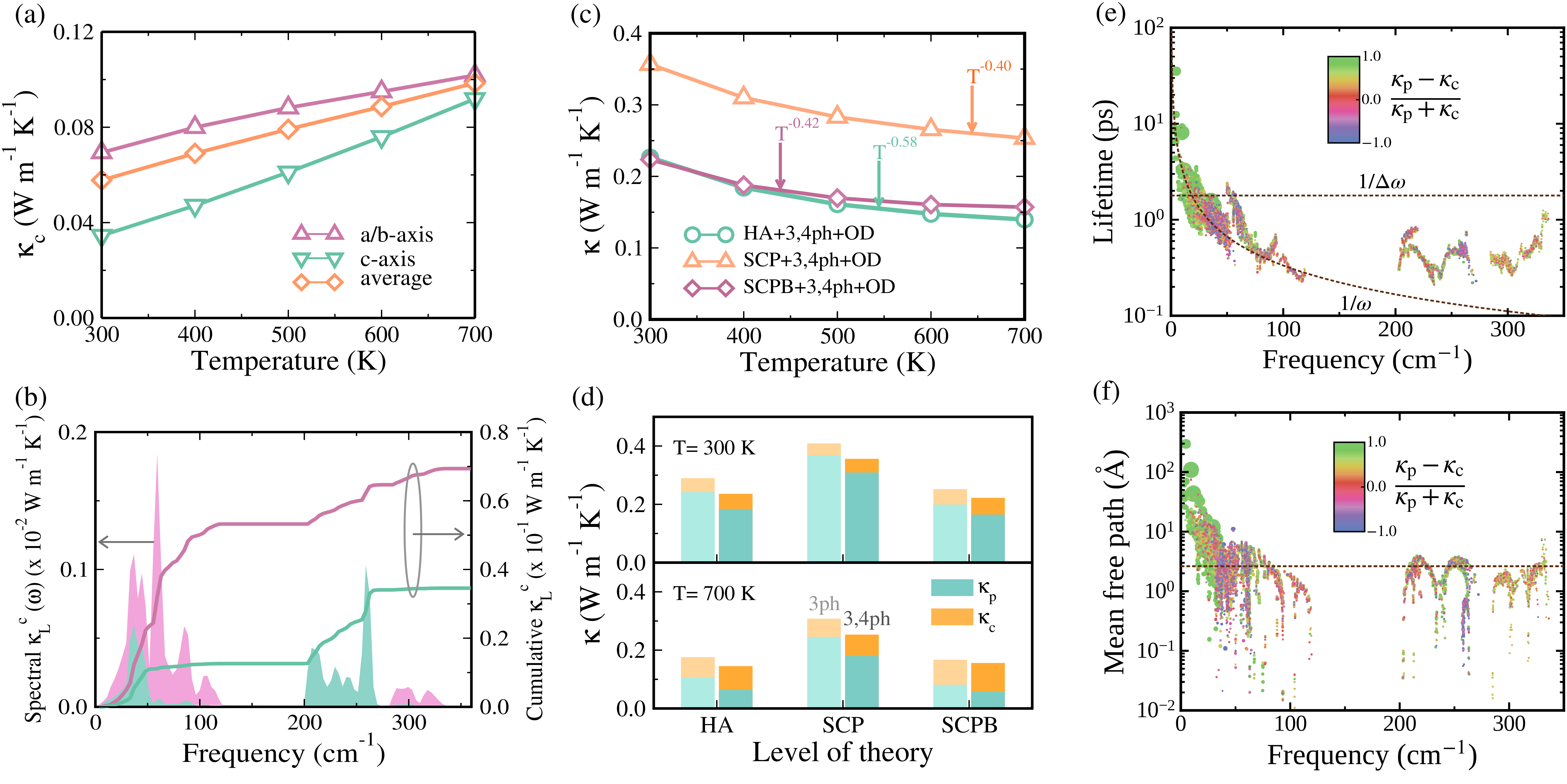}
\caption{(a) Temperature-dependent coherence's 
thermal conductivity ($\kappa_c$) within 
the SCPB+3,4ph framework. (b) Spectral (shaded, left
axis) and cumulative (line, right axis) $\kappa_c$ within 
the SCPB+3,4ph framework. (c) Temperature-dependent 
total lattice thermal conductivity 
$\kappa = \kappa_p + \kappa_c$
for CsAg$_3$S$_2$ computed
within different theoretical frameworks.
The downward arrows represent the 
fitted temperature dependence
$\kappa_p \propto T^{-\alpha}$. 
(d) Contribution of $\kappa_p$
and $\kappa_c$ to $\kappa$ at 300 K
and 700 K using different theoretical
frameworks considering 3ph and 3,4ph
scattering processes.
(e) Phonon lifetimes and (f) mean-free-paths
for CsAg$_3$S$_2$ as a function of frequency
at 300 K. The circle area is proportional
to the modal contribution to $\kappa$
and the colorbar indicates the origin
of the transport, from particle-like
(green color) to wave-like (blue color).
In (e), the horizontal line denotes the
Wigner limit in time $\tau=1/\Delta\omega$,
and the curve marks the Iofee--Regel limit
in time ($\tau=1/\omega$). In (f), the
horizontal line represents the Iofee--Regel
limit in space.}
\label{wavelikekpa}
\end{figure*}

To gain further insight into the microscopic
origin of the ultralow $\kappa_p$, we
analyze the spectral and cumulative lattice
thermal conductivities, in Fig.~\ref{kpa} (d)-(i).
Irrespective of the theoretical 
framework adopted, the cumulative conductivity 
shows that the c-axis thermal  transport
is almost entirely contributed by phonon modes
below frequency $\sim$50 cm$^{-1}$, whereas along 
the $a/b$-axis more than half of $\kappa_p$ originates
from the same frequency range and exceeds
80 \% after including modes up to $\sim$130 cm$^{-1}$. Together
with the projected phonon density of 
states, this shows that the c-axis transport
is governed by Cs-dominated acoustic and low-lying optical 
modes which explains its strong sensitivity to
weak Cs-S coupling, while Ag-related
modes between 50 and 150 cm$^{-1}$ provide
an additional in-plane contribution. The
high-frequency optical modes contribute
negligibly. The inclusion of 4ph scattering
reduces the magnitude of $\kappa_p$
without appreciably changing the frequency
distribution of heat-carrying phonons.

To further elucidate the origin of suppressed
$\kappa_p$, we examine 
the phonon group velocities ($v_g$) in
Fig.~\ref{scattering} (a)-(c). The group 
velocities are higher in the in-plane direction than
along the out-of-plane directions, which is consistent
with the bonding hierarchy and the anisotropy 
in $\kappa_p$. The acoustic branches retain the
largest $v_g$ and dominate
the particle-like transport. Importantly, the
evolution of $v_g$ across the low- and 
intermediate-frequency regimes reveals two complementary 
mechanisms for suppressing $\kappa_p$. A pronounced
dip in $v_g$ around 40-60 cm$^{-1}$ arises
from the Cs-dominated rattling modes, which flatten the
low-frequency optical branches at the avoided
crossing. Beyond this region, $v_g$ decreases
progressively across the Ag-dominated part 
of the spectrum, reflecting the heavier and
weakly bonded Ag-S framework compared to
Cu-S framework in CsCu$_3$S$_2$, which lowers the
phonon frequencies. Thus, while
the Cs-rattling primarily impedes the
low-frequency phonons, the softer Ag-S
framework extends the suppression of
$v_g$ to the heat-carrying low-lying
optical modes, indicating that the two
mechanisms work cooperatively to
reduce $\kappa_p$. The recovery of $v_g$ 
in the optical branches above
the phonon gap is consistent with
the small contribution to the cumulative
$\kappa_p$ near 200 cm$^{-1}$, although
its overall magnitude remains low.    

We next examine the phonon scattering rates
governing the phonon lifetimes and thermal 
transport in Fig.~\ref{scattering} (d)-(f). The
scattering rates, in general, increase with 
frequency, with negligible isotope scattering
compared to the anharmonic 3ph and 
4ph contributions. 3ph
scattering dominates over most of the frequency 
spectrum; however, near the avoided 
crossing region, the 3ph rates
decrease  while the 4ph rates increase sharply
and become comparable to the 3ph
contributions. The interaction between
the acoustic and low-lying optical branches
opens a small frequency gap and flattens
the dispersions, restricting certain 3ph processes
that can simultaneously satisfy energy
and momentum conservation. In contrast, the 
localized, rattling-like Cs vibrations
weakly bonded to the Ag-S framework, 
provide strong fourth-order anharmonicity 
and enhanced 4ph scattering. 
A similar behavior was reported for
Cs$_2$AgBiBr$_6$, where the 4ph scattering
rates are comparable to the 3ph rates at 300~K
\cite{Zheng24}. The fourth-order
processes therefore make a significant 
contribution to the suppression of particle-like
thermal transport in CsAg$_3$S$_2$.

To gain further insights into the underlying
scattering mechanisms, we resolve the 3ph and 4ph
scattering rates into their respective sub-processes.
The 3ph processes comprise recombination
$(\lambda + \lambda^{'} \rightarrow \lambda^{''})$ and
splitting $(\lambda \rightarrow \lambda^{'} + \lambda^{''})$.
As shown in Fig.~\ref{subprocessscattering}
(a)-(c), recombination dominates at low 
frequencies, whereas splitting grows 
rapidly with frequency and dominates 
for the high-frequency optical modes, consistent
with the larger splitting phase 
space (Fig. S7 (a)-(c)\cite{suppl}).
Notably, only the recombination channel is
suppressed in the avoided crossing 
region, indicating that the localized Cs-derived
modes restrict recombination processes and 
thereby reduce the total 3ph scattering rates.
Similarly, the 4ph processes can 
be decomposed into recombination
$(\lambda + \lambda^{'} + \lambda^{''} \rightarrow \lambda^{'''})$,
splitting $(\lambda  \rightarrow \lambda^{'} + \lambda^{''} + 
\lambda^{'''})$, and redistribution $(\lambda + \lambda^{'}
\rightarrow \lambda^{''} + \lambda^{'''})$ sub-processes.
As shown in Fig.~\ref{subprocessscattering} 
(d)-(e), the redistribution channel 
dominates throughout the spectrum  
and exhibits a pronounced enhancement near
the avoided crossing. The corresponding
4ph scattering phase space also reaches 
its maximum in this 
region [see Fig. S7 (d)-(e) \cite{suppl}],
establishing redistribution as the main
source of the enhanced fourth-order
scattering corresponding to the
localized Cs modes. Similar 
redistribution-dominated 4ph scattering is also 
reported in Cs$_2$AgBiBr$_6$ \cite{Zheng24}, while
flat rattler modes driven strong
4ph scattering was observed in 
CsCu$_4$Se$_3$ \cite{yue25}, AgCrSe$_2$ \cite{xie20} and 
Cu$_{12}$Sb$_4$S$_{13}$ \cite{Yixia20}. These results
highlight the important role of higher-order
anharmonic scattering in suppressing the
particle-like thermal transport in CsAg$_3$S$_2$.    

We further resolved the 4ph scattering rates
into Umklapp (U) and normal (N) processes 
[Fig.~\ref{subprocessscattering} (g)-(h)].
U processes violate crystal-momentum
conservation and directly provide thermal 
resistance, whereas N processes conserve
total crystal momentum and redistribute
it among the phonons without directly generating 
any thermal resistance. Both
U and N rates increase with frequency and
peak near the avoided crossing region, indicating that
the enhanced 4ph scattering in this frequency region 
arises from both channels. Additionally, the
U/N ratio provided in the inset reveals the dominance of
the U processes in the 4ph scattering rates, suggesting
that the 4ph scattering suppresses $\kappa_p$
primarily by generating thermal resistance.

To elucidate the respective effects of anharmonic
phonon renormalization and 4ph 
scattering on particle-like thermal transport, we
examine the thermal conductivity ratios
$\kappa_{3,4\mathrm{ph}}^{\mathrm{SCP}}/\kappa_{3,4\mathrm{ph}}^{\mathrm{HA}}$, $\kappa_{3,4\mathrm{ph}}^{\mathrm{SCPB}}/\kappa_{3,4\mathrm{ph}}^{\mathrm{SCP}}$,
and $\kappa_{3,4\mathrm{ph}}^{\mathrm{SCPB}}/\kappa_{3,4\mathrm{ph}}^{\mathrm{HA}}$,
for renormalization and $\kappa_{3,4\mathrm{ph}}^{\mathrm{HA}}/\kappa_{3\mathrm{ph}}^{\mathrm{HA}}$, $\kappa_{3,4\mathrm{ph}}^{\mathrm{SCP}}/\kappa_{3\mathrm{ph}}^{\mathrm{SCP}}$, and $\kappa_{3,4\mathrm{ph}}^{\mathrm{SCPB}}/\kappa_{3\mathrm{ph}}^{\mathrm{SCPB}}$
for 4ph scattering \cite{Zhao21} in
Fig. S9 of SI \cite{suppl}. 
The loop-only renormalization increases
$\kappa_p$ by  70\%  relative to
the harmonic-only treatment, whereas the bubble
correction mitigates
this effect, yielding a 10\% decrease
relative to the harmonic-only value.
Consequently, the bubble correction 
reduces $\kappa_p$ by $\sim$ 47\%
relative to the SCP theory, significantly
larger than the $\sim$18\% decrease reported for
isostructural CsCu$_3$S$_2$ \cite{yue24}
for the same level of theory. This 
indicates a considerably stronger
cubic anharmonicity in CsAg$_3$S$_2$, consistent
with its weaker bonding.
Four-phonon scattering further 
suppresses $\kappa_p$ with stongest
reduction within HA
(24 \%), followed
by SCPB (17 \%) and SCP (16 \%) at
300 K. The detailed temperature
and axis dependence of these ratios
are discussed in the SM \cite{suppl}.  

\begin{figure*}
\includegraphics[width=0.9\textwidth]{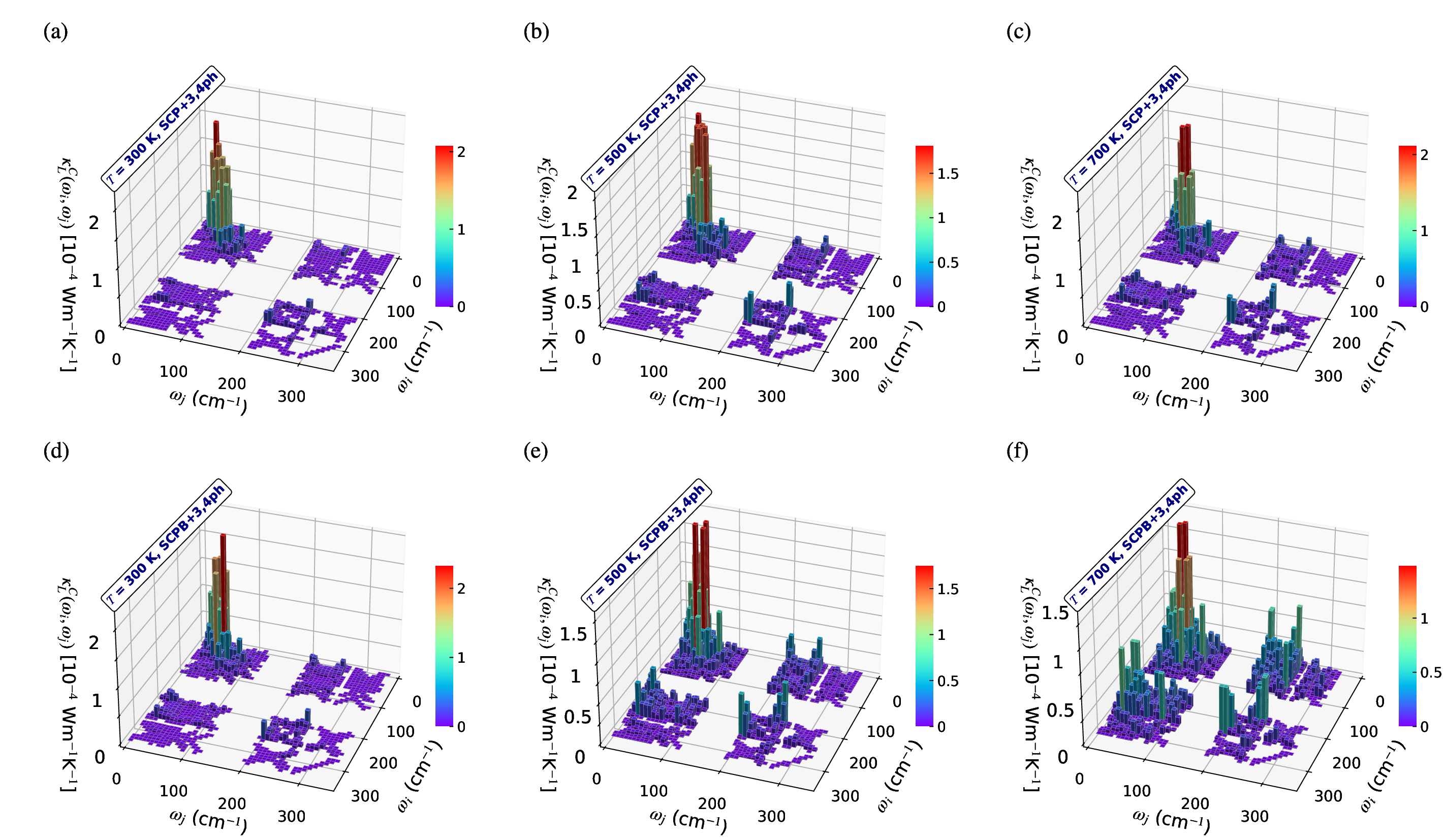}
\caption{(a) Three-dimensional visualization of modal
contributions to coherence's thermal conductivity
resolved in terms of frequencies $\omega_1$ and
$\omega_2$ at 300 K within the 
SCP+3,4ph framework. (b) Same as 
(a) but at 500 K. (c) Same as (a) but at
700 K. (d) Same as (a) but within the
SCPB+3,4ph framework. (e) Same as (d) but at 500 K.
(f) Same as (d) but at 700 K. Contributions along
the diagonal ($\omega_i = \omega_j$)
arise from the coupling between the 
quasi-degenerate phonon modes.}
\label{modalcoherence}
\end{figure*}

\subsection{Wave-like Thermal Transport}
Having established the strong suppression
of particle-like thermal transport in CsAg$_3$S$_2$, we
now turn our attention to the coherence's 
contribution ($\kappa_c$) to thermal transport. $\kappa_c$ 
originates from the off-diagonal component of the
Wigner heat-flux operator \cite{Allen93}. When the
phonon linewidth becomes comparable to the interbranch
spacing, this leads to coherent Zener-like tunneling 
between phonon modes, enabling wave-like heat
transport \cite{Simoncelli19}. This provides an 
intrinsic challenge for reducing the total thermal
conductivity. The structural features that 
suppress $\kappa_p$ through strong anharmonic scattering 
and phonon mode softening can simultaneously broaden
phonon linewidths and enhance interbranch 
coupling, thereby increasing $\kappa_c$
\cite{rodriguez26}. This compensation in CsAg$_3$S$_2$ remains 
remarkably weak. The calculated $\kappa_c$ at 
300 K is 0.057 W m$^{-1}$ K$^{-1}$ and increases 
to 0.098 W m$^{-1}$ K$^{-1}$ at 700 K
within the SCPB+3,4ph framework and remains
below 0.1 W m$^{-1}$ K$^{-1}$ for all levels of 
theoretical frameworks considered 
[see Fig.~\ref{wavelikekpa} (a) and
Figs. S10 (a)-(c) and S11 (a)-(c) in SI \cite{suppl}]. The origin of 
this suppression can be traced to the phonon 
dispersions. The strongly bonded light S atoms form
sparse, spectrally isolated, high-frequency
optical modes with large interbranch separations, thereby
reducing the coherent coupling between the 
neighbouring phonon modes.
Importantly, although the weaker Ag-S
bonding shifts the spectrum to lower
frequencies than in CsCu$_3$S$_2$, the
phonon gap and the large interbranch
separation persists, so the substitution
that suppresses $\kappa_p$ does
not compromise the spectral sparsity
that limits $\kappa_c$. Thus, while
the weakly bonded Cs sublattice and the softer
Ag-S framework suppress $\kappa_p$, the
S-dominated high-frequency manifold
independently constrains $\kappa_c$, so
that both transport channels are 
limited simultaneously by distinct features of 
hierarchical lattice dynamics

To elucidate the microscopic origin of
coherence's thermal conductivity, we further
analyze the spectral and cumulative $\kappa_c$ in
Fig.~\ref{wavelikekpa} (b) and Figs. S10 (d)-(f),
S11 (d)-(f) \cite{suppl}. The
coherence's contribution is concentrated predominantly
below \(\sim100\) cm\(^{-1}\), with most
in-plane $\kappa_c$ arising from
this frequency window. Along the out-of-plane
direction, $\kappa_c$ exhibits two distinct windows
contributing to $\kappa_c$; below
\(\sim50\) cm\(^{-1}\) and 
between \(\sim200\)–270 cm\(^{-1}\). The 
latter coincides with the lower part of the 
S-dominated optical modes, indicating a 
near-degenerate mode coupling. Beyond this
region, the large interbranch separations suppress
coherent coupling, rendering the 
high-frequency optical modes inactive in
heat transfer. Thus, while strong anharmonicity
promotes coherent transport through linewidth 
broadening, substantial $\kappa_c$ additionally requires
a dense, closely packed phonon spectrum of 
energetically accessible neighbouring modes. This 
spectral contrast is also observed in 
isostructural CsCu$_3$S$_2$ \cite{yue24} and 
Cs$_6$Re$_6$S$_8$I$_8$ \cite{xiong25}, where 
sparse optical modes contribute
weakly to wave-like thermal transport.

Combining both channels, $\kappa = \kappa_p  +\kappa_c$
we obtain an ultralow thermal conductivity of
0.222 W m$^{-1}$ K$^{-1}$ at 300 K, which further
reduces to 0.156 W m$^{-1}$ K$^{-1}$ at 700 K within
the SCPB+3,4ph framework 
[see Fig.~\ref{wavelikekpa} (c)].
Because the growing contribution from
$\kappa_c$ partially offsets the effect
of decreasing $\kappa_p$, $\kappa$ follows
a weak temperature dependence of $\sim$T$^{-0.42}$,
in contrast to the $\sim$T$^{-1.25}$ and 
$\sim$T$^{-1.04}$ dependencies of in-plane and
out-of-plane $\kappa_p$, respectively.
The room temperature values are comparable
to those of materials with simultaneously
suppressed $\kappa_p$ and $\kappa_c$, such as
such as Cs$_6$Re$_6$S$_8$I$_8$ \cite{xiong25} 
and Tl$_4$GeS$_4$ \cite{wu26}. Furthermore, $\kappa$ in
CsAg$_3$S$_2$ is nearly an order of magnitude smaller
than 1.66 W m$^{-1}$ K$^{-1}$ reported 
for isostructural CsCu$_3$S$_2$, reflecting
the combined weakening of the Ag-S framework and 
Cs-S coupling upon Cu to Ag substitution.

The SCP+3,4ph description consistently 
overestimates $\kappa$ relative to the SCPB+3,4ph
by 60.36 \% at 300 K and 62.17 \% at 700 K, highlighting
the importance of the cubic bubble correction. 
Interestingly, the HA+3,4ph and
SCPB+3,4ph approaches yield nearly 
identical $\kappa$ at 300 K despite their substantially
different $\kappa_p$, owing to the 
enhanced $\kappa_c$ within the SCPB treatment, which 
compensates for its lower $\kappa_p$. This
agreement disappears at elevated temperatures as
$\kappa_c$ acquires a stronger temperature dependence
within the SCPB+3,4ph framework. The inclusion of
four-phonon scattering further suppresses $\kappa$ for 
all renormalization schemes, although the magnitude 
of this reduction decreases at elevated 
temperatures. This can be attributed to the 
growing $\kappa_c$ , which partially
compensates for the
reduction in $\kappa_p$.

Next, we quantify and analyze the individual
contributions of $\kappa_p$ and $\kappa_c$ to 
$\kappa$ in Fig.~\ref{wavelikekpa} (e)
and Table S3 in SM \cite{suppl}. The
relative contributions reveal a 
pronounced temperature-driven crossover
in CsAg$_3$S$_2$. At 300 K, particle-like
transport dominates within the SCPB+3,4ph
framework, with $\kappa_p$ accounting for 
approximately 74 \% of the total $\kappa$. With
increasing temperature, $\kappa_p$ decreases 
rapidly, whereas $\kappa_c$ increases. At 700 K, the
coherence channel contributes 62.82 \% surpassing 
$\kappa_p$. A similar
particle-to-wave crossover has been reported in 
strongly anharmonic materials such
as Cu$_3$BiS$_3$ \cite{yue25a}, Tl$_2$AgI$_3$ \cite{pathak26},
AgTlI$_2$ \cite{ju25}. In contrast, $\kappa_p$
remains dominant in CsCu$_3$S$_2$ up
to 900 K \cite{yue24}, indicating that
the crossover is enabled by the stronger
suppression of $\kappa_p$ upon Cu to
Ag substitution. The important distinction 
in CsAg$_3$S$_2$ is that the crossover results
primarily from the continued suppression
of $\kappa_p$ by weakly bound Cs atoms and softer
Ag-S framework, rather than from the emergence of
large absolute coherence conductivity, as $\kappa_c$ 
remains below 0.1 W m$^{-1}$ K$^{-1}$. Therefore, this
crossover reflects the progressive collapse
of particle-like channel against a comparatively weak
coherence background. Therefore, CsAg$_3$S$_2$ 
provides a platform in which the
dominant transport mechanism changes with 
temperature while both
channels remain strongly suppressed.

To further elucidate the origin of the
suppressed coherence channel and ultralow thermal
conductivity in CsAg$_3$S$_2$, we analyze
the frequency-dependent lifetime and mean free 
path in Fig.~\ref{wavelikekpa} (e) and
(f). Here, the area of each circle is proportional
to each mode's contribution to $\kappa$, and
the color bar represents the origin of 
the contribution. The green color represents the
particle-like contribution, whereas the blue 
color corresponds to the wave-like tunneling 
channel. The intermediate colors represent a
regime where both transport channels coexist and 
contribute to the heat conduction. Modes with 
lifetime above the Wigner limit in time, defined 
as \(\tau_W=1/\Delta\omega_{\rm avg}\) \cite{Simoncelli22}, retain 
predominantly particle-like character, 
whereas modes with \(\tau\lesssim\tau_W\) progressively 
acquire the wave-like character. As shown in
Fig.~\ref{wavelikekpa}(e), the acoustic and 
nearby low-lying optical modes largely remain 
above \(\tau_W\) and dominate the thermal
transport. The high-frequency modes increasingly
fall below this limit, but contribute only 
weakly to overall thermal conductivity, consistent 
with the spectral $\kappa_c$ in 
Fig.~\ref{wavelikekpa} (b). An analogous
picture emerges in the frequency-mean-free-path
domain [Fig.~\ref{wavelikekpa}(f)], where
the high-frequency modes fall below the
Ioffe-Regel limit in space \cite{Allen93, allen99}, at
which the mean-free-path becomes comparable to
the average interatomic spacing, indicating a
predominantly wave-like character of these modes.
The contribution to $\kappa$, however is small
owing to the large separation of the
high-frequency branches. Thus, the loss of
quasi-particle character does not
ensure a large coherent contribution.

Fig.~\ref{modalcoherence} (a)–(c) and (d)–(f)
resolve $\kappa_c$ in terms of
the frequencies $\omega_1$ and $\omega_2$ of the two 
coupled modes within the SCP+3,4ph
and SCPB+3,4ph frameworks, respectively.
At 300 K, $\kappa_c$ originates mainly from
coupling between the quasi-degenerate phonon modes 
below 100 cm$^{-1}$. This can be attributed 
to the strong 4ph scattering rates, and hence 
broader phonon linewidths of highly localized 
phonon modes associated with the
rattling-like vibrations of weakly bound
Cs atoms. In contrast, the high-frequency
modes above $\sim$100 cm$^{-1}$ have minimal
contribution, consistent with the cumulative
$\kappa_c$ analysis related to Fig.~\ref{wavelikekpa} (b). 
At higher temperatures, increased scattering
broadens the linewidths and promotes
coupling between the phonon 
modes of increasingly dissimilar
dissimilar frequencies. A comparison of the
two frameworks shows that the enhancement is 
stronger for the SCPB+3,4ph framework than for the
SCP+3,4ph framework. Accordingly, at 700 K, the 
contribution from the high-frequency phonons 
with larger frequency differences has a finite 
contribution to coherence's thermal transport. 
This behavior differs from 
Cs$_2$AgBiBr$_6$ \cite{Zheng24}, where the high-frequency 
modes contribute significantly to
$\kappa_c$. This phenomenon resembles
 the behavior observed in CsCu$_3$S$_2$
 \cite{yue24},  Cs$_6$Re$_6$S$_8$I$_8$ \cite{xiong25}, 
and Tl$_4$SiS$_4$ \cite{wu26}, where the high-frequency
modes do not actively participate in the 
wave-like thermal transport. Together,
these results show that the weakly bound
Cs atoms and the softer Ag-S framework drive
low $\kappa_p$, while the sparse S-derived
optical branches, preserved upon substitution,
keep $\kappa_c$ small, leading to the
simultaneous suppression of both
transport channels.

\subsection{Electronic Structure and Thermoelectric Transport}
The obtained ultralow thermal conductivity 
makes CsAg$_3$S$_2$ an interesting prospect for
thermoelectric applications. Therefore, to complement
the analysis of weak thermal transport, we further
examined the electronic structure and electrical 
transport properties to assess the overall thermoelectric
performance. The calculated electronic band 
structure [see Fig. S14 \cite{suppl}] reveals a semiconducting
nature with an indirect band gap of 2.2 eV. Additionally, the
multi-valley nature in the conduction band 
region, in conjunction with the multi-valley convergence 
in the valence band region around point A, further 
provides a suitable electronic environment for favorable 
electrical transport properties \cite{pei11}. Furthermore, we 
note a distinct orbital character in the 
electronic band structure. The valence band 
maximum (VBM) is predominantly derived from the
Ag 4$d$ and S 3$p$ states, whereas the conduction
band minimum (CBM) is mainly composed of
Ag 5$p$ and S 3$p$ states. This indicates that
the electronic states near the band edges are primarily
governed by the Ag-S framework, with 
dominant $p$-$d$ hybridization in the CBM region 
and $p$-$p$ hybridization in the VBM region
which is consistent with the antibonding Ag-S states
near the Fermi level identified in the COHP analysis.

The carrier-dependent transport
coefficients (Fig. S16 \cite{suppl}), obtained
within the MRTA, exhibit
the expected competition between the Seebeck 
coefficient ($S$) and electrical conductivity ($\sigma$). The
Seebeck coefficient increases with temperature but 
decreases with the carrier concentration. In
contrast, electrical conductivity increases
with carrier concentration and decreases with
temperature. The electronic thermal conductivity follows
the trend of electrical conductivity
closely, consistent with the Wiedemann-Franz law. The
relatively large band gap further suppresses
bipolar transport throughout the investigated 
temperature range. Despite the favorable $S$, the
comparatively low $\sigma$ limits the power
factor. Combining S and $\sigma$, we obtained the
highest power factor values of
8.07 $\mu\mathrm{W\,cm^{-1}\, K^{-2}}$ and
3.17 $\mu\mathrm{W\,cm^{-1}\, K^{-2}}$ for n- and
p-type doping, respectively, at 700 K.  

\begin{figure}
\includegraphics[width=0.9\columnwidth]{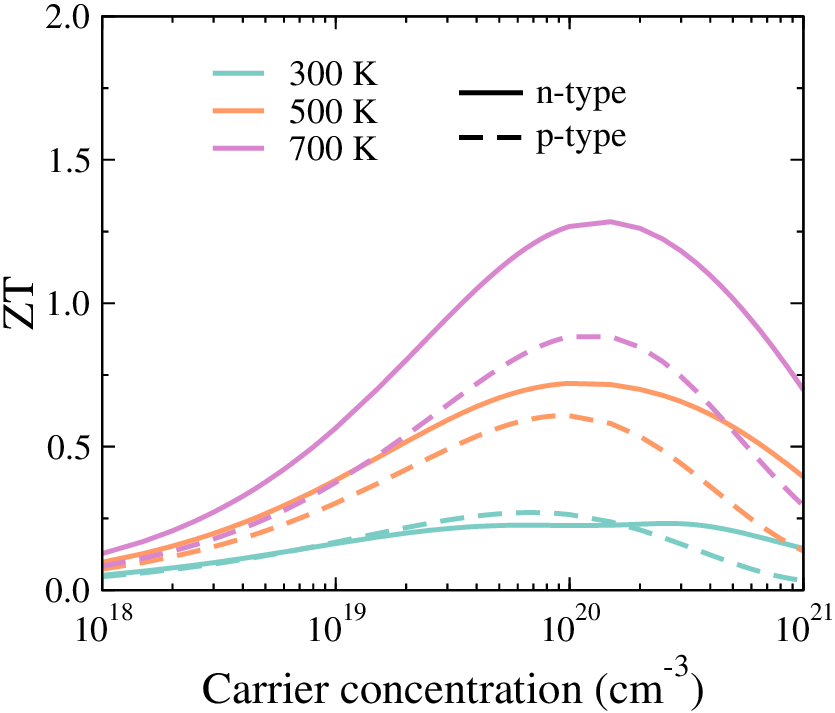}
\caption{Thermoelectric figure of merit as a
function of carrier concentration at various temperatures 
for $n$- and $p$-type doping.}
\label{zt}
\end{figure}

The resulting thermoelectric figure
of merit ($ZT$) [see Fig.~\ref{zt}], determined
by the $PF$ and the ultralow lattice thermal
conductivity, displays a characteristic
non-monotonic dependence on carrier concentration.
$ZT$ increases initially with carrier concentration for
either type of doping to reach a maximum value 
at an optimum carrier concentration, and subsequently decreases
at higher carrier concentrations. Furthermore, the
calculated ZT increases with temperature, reaching
maximum values of approximately 1.28 
and 0.88 at 700 K for n-type and p-type
doping, respectively, corresponding to a carrier
concentration of $1.5 \times 10^{20}$
cm$^{-3}$. Altogether, these results indicate
that, although the electrical transport properties 
contribute to the thermoelectric 
response, the ultralow lattice thermal 
conductivity plays a key role in enabling 
respectable $ZT$ values at elevated temperatures.

\section{Conclusion}
In summary, we investigated the structural, 
vibrational, and lattice thermal transport 
properties of trigonal CsAg$_3$S$_2$ using 
first-principles calculations in conjunction with
self-consistent phonon theory, higher-order
anharmonic scattering, and unified theory of thermal 
transport. Although CsAg$_3$S$_2$ has so far
been reported in the monoclinic symmetry, the
trigonal phase is energetically accessible,
dynamically, mechanically, and structurally
stable up to 700 K. The bonding analysis suggests
the presence of a unique bonding hierarchy consisting
of covalently bonded [Ag$_3$S$_2$]$^{-}$ 
units and a loosely bonded Cs$^{+}$ sublattice.
Compared to isostructural CsCu$_3$S$_2$, the
Cu to Ag substitution weakens both the covalent
framework and the Cs-S interactions, giving
rise to an exceptionally low $\kappa$
of 0.222 W m$^{-1}$ K$^{-1}$ at 300 K and
0.16 W m$^{-1}$ K$^{-1}$ at 700 K with a
weak temperature dependence of
$\sim$T$^{-0.42}$. The localized Cs modes
enhance anharmonic scattering and shorten
phonon lifetimes, and the heavier, softer
Ag-S framework reduces phonon 
group velocities, together suppressing
particle-like transport, while the
spectrally sparse and well-separated
S-derived optical branches restrict
coherent phonon coupling, leading
to the simultaneous suppression of both
transport channels. With increasing 
temperature, the coherent contribution 
rises from about 26 \% to 63 \%, leading
to a crossover from particle-like thermal
transport to wave-like tunneling.
This crossover is not observed in CsCu$_3$S$_2$
and arises primarily from the decrease of
$\kappa_p$ rather than from a large increase
in $\kappa_c$. Combined with the computed
electrical transport parameters, this
yields a $ZT$ of approximately 1.28 at 
700 K for n-type doping.   

Beyond establishing ultralow thermal conductivity
and clarifying the distinct roles of particle-
and wave-like transport channels in the thermal
transport of trigonal CsAg$_3$S$_2$, our results
show that modifying the covalent framework in
low thermal conductivity materials featuring
bonding hierarchy structural motif can provide
an additional route to suppress $\kappa_p$
without disturbing the mechanisms limiting $\kappa_c$.
In CsAg$_3$S$_2$, the heavier and softer Ag-S
framework suppresses the phonon group
velocities, while the accompanying weakening
of Cs-S coupling reduces the phonon lifetimes
by promoting the rattler-driven anharmonic
scattering. Together with the sparse S-derived
optical modes limiting the coherent coupling,
these mechanisms provide a route for simultaneous
suppression of both heat transport channels.
More broadly, systematic modification of the
covalent framework across the AM$_3$Ch$_2$ family
and related bonding hierarchy materials
provide a natural platform to examine its
generality. Since the framework modification
also tunes the coupling of the ionic
sublattice, it may offer a route to modulate the relative
contribution of $\kappa_p$ and $\kappa_c$.
We believe this study will motivate further
investigations to test this strategy for design 
and discovery of ultralow thermal
conductivity materials.

\begin{acknowledgements}
 
We would like to thank Md. Zeeshan for the useful discussions. 
V.G. is thankful to BUKRP00034052 for the
financial assistance. B.K.M. acknowledges the
funding support from the SERB, DST (CRG/2022/000178). 
The calculations
were performed using TEJAS cluster at the Department of Physics, 
IIT Delhi and the PARAM Rudra, a national supercomputing facility
at the Inter-University Accelerator Centre (IUAC), New Delhi.

\end{acknowledgements}

\bibliography{csags}

\end{document}